\documentclass[12pt]{article}

\usepackage{graphicx}
\usepackage[body={17.5cm, 22cm},right=2.2cm]{geometry}
\usepackage{amssymb}
\usepackage{amsmath}
\usepackage{epsfig}
\usepackage{psfrag}

\def\const{\mbox{const}}
\def\e{{\rm e}}

\def\d{\partial}
\def\l{\left(}
\def\r{\right)}

\newcommand{\be}{\begin{equation}}
\newcommand{\ee}{\end{equation}}
\newcommand{\ba}{\begin{align}}
\newcommand{\ea}{\end{align}}
\newcommand{\bg}{\begin{gather}}
\newcommand{\eg}{\end{gather}}
\newcommand{\bseq}{\begin{subequations}}
\newcommand{\eseq}{\end{subequations}}

\renewcommand{\ln}{\mathop{\rm ln}\nolimits}

\begin{document}

\setcounter{page}{0}
\begin{titlepage}
\begin{center}

\vspace{1.cm}

{\Large
\bf{
Bumpy black holes from spontaneous Lorentz violation
}}\\[1cm]

{\large Sergei Dubovsky$^{\rm a,b}$, Peter Tinyakov$^{\rm c,b}$, Matias Zaldarriaga$^{\rm a,d}$
\\[0.5cm]

{\small \textit{$^{\rm a}$ Jefferson Physical Laboratory, Harvard University, Cambridge, MA 02138, USA}}

\vspace{.2cm}
{\small \textit{$^{\rm b}$ Institute for Nuclear Research of the Russian Academy of Sciences, \\
        60th October Anniversary Prospect, 7a, 117312 Moscow, Russia}}

\vspace{.2cm}
{\small \textit{$^{\rm c}$ Service de Physique Th\' eorique, Universit\' e Libre de Bruxelles,\\
CP225, blv. du Triomphe, B-1050 Bruxelles, Belgium}}

\vspace{.2cm}
{\small \textit{$^{\rm d}$ Center for Astrophysics, Harvard University, MA 02138, USA}}
}

\end{center}
\begin{abstract}
We consider black holes in Lorentz violating theories of massive
gravity. We argue that in these theories black hole solutions are no
longer universal and exhibit a large number of hairs. If they exist,
these hairs probe the singularity inside the black hole providing a
window into quantum gravity.  The existence of these hairs can be
tested by future gravitational wave observatories. We generically
expect that the effects we discuss will be larger for the more massive
black holes. In the simplest models the strength of the
hairs is controlled by the same parameter that sets the mass of the
graviton (tensor modes). Then the upper limit on this mass coming from
the inferred gravitational radiation emitted by binary pulsars
implies that hairs are likely to be suppressed for almost the entire
mass range of the super-massive black holes in the centers of galaxies.
\end{abstract}
\noindent
{\small 
} \vfill
\end{titlepage}
\setcounter{footnote}{0}
\tableofcontents
\section{Introduction and summary}
\label{intro}
Gravity remains the most mysterious force in nature, affecting
properties of space and time at the most fundamental level. Large
quantum fluctuations of the metric at the Planck scale indicate that
the very basic principles of quantum field theory, such as locality,
are likely to be drastically modified in quantum gravity.
Unfortunately, there is little hope to directly probe gravity in this
regime.

The existence of black holes provides an alternative window to explore
non-perturbative gravitational dynamics at energy densities well below
the Planck scale.  Indeed, for a long time black holes have been a
principal ``theoretical laboratory" for quantum gravity.  Many of the
advances in string theory resulted from attacking the fundamental
puzzles of black hole thermodynamics. Moreover, black holes hide a
singularity in their interiors that probes the quantum gravity
regime. Unfortunately, ``cosmic censorship" appears to hide from our
view what is happening at the singularity.

In practice  General Relativity (GR) is routinely used to
describe gravitational phenomena that span a wide range of scales from
the Solar system to the entire Universe. GR has survived many
precision tests in the Solar system and has successfully predicted the
emission of gravitational waves by binary pulsars (see, eg.,
\cite{Will:2005va}). However, most of the existing tests concern the
weak field regime. The tests of GR in the strong field limit, i.e.,
when the non-linearities of the Einstein equations play an essential
role, are more difficult to obtain.

One such case is the dynamics of the Universe as a whole. The
cosmological model based on GR is confirmed by observations with an
ever-growing precision \cite{Spergel:2006hy} up to the nucleosynthesis
epoch.  This model requires, however, that the present Universe is
dominated by a dark energy and dark matter of as yet unknown
origin. It is not clear, therefore, to what extent the above agreement
can be considered as a confirmation of GR itself.

Another possibility to study non-linear effects of gravity is
provided by the astrophysical black holes that are expected to be a
perfect laboratory for quantitative tests of GR in the strong field
limit. It is conceivable that in the near future the validity of the
Schwarzschild or Kerr metric around astrophysical black holes could be
tested with high precision through a variety of astronomical
observations.

There exists two techniques which may allow to reconstruct the metric
of the black hole close to the horizon. The first one consists of
observing the electromagnetic radiation which comes from the innermost
region of the accretion disk of a black hole and encodes information
about the space-time structure of that region. Extracting this
information requires detailed modeling of the accretion disk in order
to disentangle the physical effects which depend on the structure of
the metric (such as strong gravitational lensing, large redshifts and
time delays) from other unknowns such as the details of the physical
state of the gas in the disk, the accretion rate or the mechanism
responsible for the emission of the electromagnetic radiation
(eg. \cite{McClintock:2006xd}). The GR effects influence the emission
coming from different parts of the disk in different ways, sometimes
leading to ``easily" identifiable features in the light curve (see,
e.g., \cite{Broderick:2005at}). For a recent review of astrophysical
black holes and the various observational techniques to characterize
them see \cite{Narayan:2005ie}.

Another technique being developed for future gravitational wave
observatories relies on the detailed study of the time dependence of
the emitted radiation during the inspiral and merger phases of binary
systems involving at least one black hole. For the purpose of testing
GR the most promising candidates are perhaps the so-called extreme
mass ratio inspirals (EMRI), that is compact stellar mass objects
captured by supermassive black holes ($\sim 10^6 \ M_\odot$) in
galactic nuclei. Because of the large mass ratio the small compact
object is to a very good approximation a test particle orbiting around
a black hole. LISA is expected to detect about a hundred of such EMRIs
per year. Because of the small mass ratio the inspiral should be
observable by LISA for years, during roughly $10^5$ orbits. The study
of such events could lead to strong constraints on the multipole
moments of the space-time around the black hole and thus provide a
precision test of the black hole metric.

A point important for the purpose of this paper is that the multipole
moments of the black hole metric are very sensitive to potential
deviations from GR. Indeed, a striking property of black holes in GR
is the absence of
``hairs"~\cite{Bekenstein:1971hc}-\cite{Price:1971fb2}, namely no
matter what the shape of the collapsing object is, all multipole
moments of the resulting black hole are determined just by its total
mass and angular momentum. In fact, this no-hair property is one of
the main features which distinguishes black holes from ordinary
massive non-radiating objects.  With LISA one expects to be sensitive
to 6-7 lowest black hole multipoles with a precision at the level of a
few percents, and thus be able to obtain a quantitative verification
of the universality of the black hole metric (the absence of ``bumps")
\cite{Hughes:2006pm,Collins:2004ex}.

Given the observational promise and the considerable efforts put to
measure the detailed properties of the astrophysical black holes, one
may wonder what are the benchmark theoretical models which provide
predictions for these observations different from those of GR. In
particular, what would be the implications of the black hole bumps, if
they were to be discovered?

Naively, one might expect that a black hole is the most natural place
to test alternative theories of gravity. 
However, the actual situation is more complicated. One of the problems
is that the space-time curvature around astrophysical black holes is
very small.  Consequently, black hole observations have practically no
chance to discover short distance modifications of gravity such as
those induced by higher dimensional operators in the gravitational
action, or those due to large extra dimensions.  Indeed, the sizes and
the curvature radii of the astrophysical black holes are at least of
the order of few kilometers (for the stellar mass black holes), while
the existing short-distance tests of the gravitational force do not
find any deviations from GR up to distances as short as a fraction of
a millimeter \cite{Smullin:2005iv,Adelberger:2006dh}. As a result,
even the most extreme scenarios with the ultra low quantum gravity
scale \cite{Dvali:2001gx} do not lead to measurable changes in the
properties of astrophysical black holes due to UV effects in the range
of parameters where they are compatible with the direct gravity
tests. In Appendix~\ref{UV} we make this argument more
quantitative.\footnote{ Note that it was suggested
\cite{Tanaka:2002rb,Emparan:2002jp} that in the Randall--Sundrum model
(that can also be thought of as a modification of gravity at short
scales) the evaporation rate of the black holes localized on our brane
can be significantly enhanced due to the presence of the continuum
spectrum of the light Kaluza--Klein modes.  This may lead to the rapid
evaporation of the astrophysical black holes with masses of order few
Solar masses. This proposal still remains somewhat controversial
\cite{Fitzpatrick:2006cd}, but even if true it does not predict
anything new for the observations of the space time around
astrophysical black holes, it just shortens their lifetime.}

As an alternative to the short distance effects, one is naturally lead
to models that modify gravity at large distances. Recently, there has
been a revival of interest in long-distance modifications of gravity
which was to a large extent motivated by the observation of the
accelerated expansion of the Universe. Though no compelling
alternative to the simplest $\Lambda$CDM scenario has emerged so far,
these efforts resulted in a much clearer theoretical understanding of
the possible models, their characteristic features and potential
observational signatures.

To find the most promising class of theories that could be tested with
black hole observations, let us start with the brief overview of
the modified gravity theories.  Probably the best studied class
of long distance modifications of gravity are scalar tensor theories
of the Brans--Dicke type. The so called $f(R)$ modifications of gravity
also belong to this category in their simplest
versions~\cite{Carroll:2003wy}.  Here the long-distance effects are
due to the presence of a new light scalar degree of freedom.  As we
will discuss shortly, the ``no-hair" theorems imply that the study of
black holes properties is not a promising way to constrain these
models.

Another class of theories includes Lorentz invariant Fierz--Pauli
model of massive gravity \cite{Pauli:1939xp} and brane world
constructions where the four-dimensional graviton mass is replaced by
a resonance with a finite width that is due to the escape of gravitons
into extra dimensions
\cite{Gregory:2000jc,Dvali:2000hr,Dvali:2000xg}. The common theme in
the study of these models is the dynamics of the longitudinal graviton
polarization, which typically leads to strong coupling (and, as a
result the loss of predictability) at an unacceptably low energy scale,
and/or to the appearance of ghosts around curved backgrounds
\cite{vanDam:1970vg}-\cite{Luty:2003vm}.

A notable exception is the five-dimensional Dvali--Gabadadze--Porrati
(DGP) brane world model where non-linear effects provide an extra
contribution to the kinetic term of the longitudinal mode of the right
sign which prevents strong coupling in the vicinity of the sources
\cite{Nicolis:2004qq} (unfortunately, this contribution has a wrong
sign for the cosmologically most interesting self-accelerating branch,
and the perturbative analysis reveals a ghost in the spectrum of
linear perturbations around this branch
\cite{Luty:2003vm,Gorbunov:2005zk,Charmousis:2006pn}).

For many purposes the DGP model can be thought of as a very peculiar
scalar-tensor theory where the derivative scalar self-interaction
results in the ``chameleon" or self-shielding behavior near massive
sources (cf. \cite{Khoury:2003aq}). This shielding is crucial for any
such theory not to be already ruled out by the Solar system tests, in
particular, by the deflection of light measurements. It results in
interesting non-linear effects at short enough distances from a
massive source giving rise, for instance, to a small anomalous
precession of the Moon perihelion. This effect is potentially
observable by the next generation of the Lunar ranging experiments
\cite{Dvali:2002vf,Lue:2002sw}.  Yet another striking result of the
non-linearities is the possibility for the superluminal propagation in
certain backgrounds \cite{Adams:2006sv}.

Finally, there exists a family of models which may be regarded as the
``Higgs phases'' of gravity in which Lorentz invariance is
spontaneously broken by condensates of scalar fields. The breaking of
Lorentz invariance that differentiates these models from the ones
discussed previously is essential to avoid the problems of strong
coupling and ghosts that plague the Lorentz invariant models
\cite{Arkani-Hamed:2003uy,Rubakov:2004eb,Dubovsky:2004sg}.  It also
allows these models to avoid the constraints coming from the
deflection of light without invoking the non-linearities. Examples of
such models are the so-called ``ghost condensate" model
\cite{Arkani-Hamed:2003uy}, as well as more general theories of
Lorentz-violating massive gravity
\cite{Dubovsky:2004sg,Dubovsky:2004ud}.  A closely related class of
models with non-trivial vacuum expectation values of the vector fields
is represented by the Einstein aether/gauged ghost condensate models
\cite{Jacobson:2000xp,Gripaios:2004ms,Cheng:2006us}. The relativistic
MOND theories were also shown to belong to this
category~\cite{Zlosnik:2006sb}.

To finish this brief survey of the infrared modifications of gravity,
it is worth noting that in spite of the considerable progress in
constructing consistent low energy effective theories that modify
gravity at long distances, none of these models (neither Lorentz
invariant nor Lorentz violating) have so far been derived from a
consistent microscopic theory. Moreover, many properties of these
theories (in particular, those related to the black hole
thermodynamics discussed later in the current paper) strongly suggest
that if such a microscopic theory exists it is likely to be very
different from string theory --- the most successful candidate for a
theory of quantum gravity --- at least in its regimes studied so far.

Which of the above classes of models, if any, are most likely to give
alternative predictions for observations of astrophysical black hole?
We have already stressed that the black hole ``no-hair" theorems
provide a very clean set of observables sensitive to the new physics
--- deviations of the black hole multipole moments from their
universal GR values. However, quite generically, these very theorems
prevent new physics from affecting the black hole metric.

To illustrate the origin of the problem let us consider a generic
model of the Brans--Dicke type, i.e., let us assume that in addition to
the metric there exists a light scalar field which by Lorentz
invariance should be coupled to the trace of the energy-momentum
tensor. Such a field provides an extra contribution to the Newtonian
1/r potential between non-relativistic sources. However, it does not
affect the deflection of light in the gravitational field of the
Sun. Consequently, the existence of such a field would lead to a
discrepancy between the values of the Solar mass deduced from the
analysis of the planetary motion and from the deflection of
light. This gives rise to stringent constraints on the strength of the
scalar force \cite{Will:2005va}.

However, if we were unlucky to have a black hole in the center of the
Solar system we would never be able to obtain such bounds. The no-hair
theorems state that the black hole horizon is not able to support a
non-zero static profile of the scalar field. Consequently, there would
be no extra force due to the scalar field and no discrepancy between
the planetary motion and the deflection of light. Thus, in the case of
the Brans--Dicke type models black holes turn out to be the worst (in
fact, hopeless) place to distinguish the conventional Einstein gravity
from a modified theory.

This example, in spite of its simplicity, actually correctly captures
the nature of the obstacles for constructing models with a modified
black hole metric. Also, it suggests that instead of trying to find a
model where the black hole metric is just slightly modified as
compared to the GR predictions, a better strategy may be to find a way
to avoid the black hole no-hair theorems altogether, so that the
higher multipole moments are not universal.

We will review in some detail the physics of the no-hair theorems in
section~\ref{nohair}, but already at the intuitive level it is clear
that these results follow from the very generic properties of the
gravitational horizons. This suggests that the best way to violate the
no-hair theorems is to consider theories with spontaneous Lorentz
breaking, where the causal structure can be modified as compared to
the standard case.

As we will explain now, this intuition can be made precise, and very
general thermodynamical considerations strongly suggest that black
holes {\it must} have hairs if Lorentz invariance is spontaneously
broken~\cite{Dubovsky:2006vk}. Recall that the way the conventional
laws of thermodynamics are recovered in the presence of black holes in
GR is truly remarkable. Indeed, one may worry that the entropy can be
lost behind the black hole horizons invalidating the second law of
thermodynamics. However, as first suggested by Bekenstein, it is
natural to assign to black holes an entropy proportional to the
horizon area. With this assignment the net entropy of a black hole and
the outer region never decreases and the second law of thermodynamics
is saved.

For this proposal to be self-consistent black holes need to have
temperature $T_H$ related to the to the energy (mass) $M$ and the
Bekenstein entropy $S_B$ in the usual way,
\[
dM=T_HdS_B\;.
\]
This is indeed true in GR with $T_H$ being the Hawking temperature of
the black hole.

To see how the black hole thermodynamics is modified in the presence of
the spontaneous Lorentz breaking, note that in this case different
species propagate with different maximum velocities $v$ even in flat
space \cite{Coleman:1998ti}. Observationally, there are extremely
tight bounds on the differences in maximum velocities for the Standard
Model fields \cite{Colladay:1998fq}. However, the experimental
constraints are easily satisfied if the hidden sector where the
Lorentz-breaking condensate develops does not have direct couplings to
the Standard Model fields, apart from those generated by graviton
loops. For the purpose of the argument below the absolute magnitude of these
coupling is irrelevant.

Progress achieved in recent years in understanding the gravitational
dynamics in the presence of spontaneous Lorentz violation made it
possible to study the consequences of the velocity differences in
curved space as well, and in particular in a black hole
background. The main result of these studies is very simple: the
effective metric describing propagation of the field with $v\neq 1$ in
the Schwarzschild background has the Schwarzschild form with a
different value of the black hole mass.  As one could have expected,
the black hole horizon appears larger for subluminal particles and
smaller for superluminal ones.  As a consequence, the temperature of
the Hawking radiation is not universal any longer; ``slow" fields are
radiated with lower temperature than the ``fast" fields.

\begin{figure}[t!]
\begin{center}
\includegraphics[width=6cm]{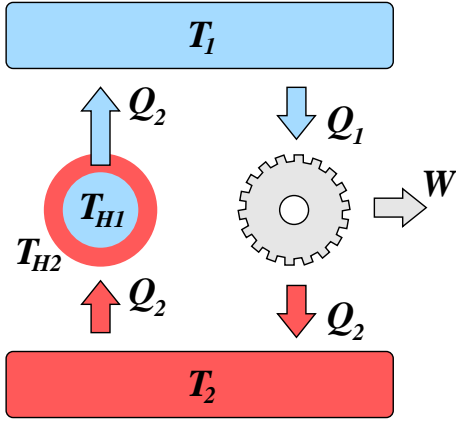}
\caption{\label{figperpetuum} \small \it In the presence of the spontaneous Lorentz breaking black holes can have different temperatures for different fields. This allows to perform thermodynamic transformations whose net effect is the transfer of heat $Q_2$ from a cold reservoir at temperature $T_2$ to a hotter one at temperature $T_1$ {\em (left)}. 
Then one can close a cycle by feeding heat $Q_1$ at the higher temperature
$T_1$ into a machine that produces work $W$ and as a byproduct releases
heat $Q_2$ at the lower temperature $T_2$ {\em (right)}. 
The net effect of the cycle is the conversion of heat into mechanical work.}
\end{center}
\end{figure}

This makes it impossible to define consistently the black hole entropy
as being determined just by its mass and the angular momentum. Indeed,
in the presence of at least two fields with different propagation
velocities, it is straightforward to provide examples of processes
such that the black hole mass and the angular momentum remain constant
while the entropy outside decreases. One example of such a process
relies on the Hawking radiation~\cite{Dubovsky:2006vk} (see
Fig.~\ref{figperpetuum}); another is a generalization of the
classical Penrose process of the energy extraction from a rotating
black hole~\cite{Eling:2007qd}.

The second law of thermodynamics follows from very basic principles of
quantum theory, such as unitarity, so in order to have a chance of
being derived from a consistent microscopic theory Lorentz violating
models have to provide a way to restore the validity of the second law
in the presence of black holes. The processes described in
Refs.~\cite{Dubovsky:2006vk,Eling:2007qd} which reduce the entropy
{\it outside} of the horizon also change the state {\it inside} the
black hole. Consequently, a contradiction with the second law can be
avoided provided this change is observable from outside. In other
words, black holes should have hairs on top of the mass and angular
momentum, which allow an observer to ``monitor" their interior state,
just like it is possible (at least in principle) for ordinary stars.

As we explain below, this indeed happens quite generally in the Higgs
phases of gravity.  Namely, a striking property of the
Lorentz-violating models with a massive graviton is the presence of
the instantaneous gravitational interactions. It is relatively easy to
understand their origin. Already in the conventional GR the graviton
propagator in non-covariant gauges (for instance, in the Newtonian
gauge) contains pieces that give rise to the static 1/r potential and
appear to be instantaneous. Of course, there are no physical
instantaneous interactions in GR; in the non-covariant gauges this
comes out as a result of the subtle cancelations between different
parts of the graviton propagator. In the Higgs phase these
cancelations are no longer exact, and physical instantaneous forces
are present. Spontaneous breaking of Lorentz invariance introduces a
preferred time in the Higgs phase and in this way the causality
paradoxes usually associated with the superluminal propagation are
avoided.
 
Given the presence of instantaneous interactions it should not be a
big surprise that black holes have an infinite amount of hairs/bumps.
What is interesting, is that the above thermodynamical argument
strongly suggests that this should be a property of all
Lorentz-violating models, excluding the ``benign" possibility that 
there exists a finite universal maximum propagating velocity (for
instance, if all fields propagate subluminally).

Apart from the Lorentz violating models, the brane world DGP model was
also found to possess a superluminal mode as a result of the
non-linear dynamics \cite{Adams:2006sv}. Its propagation velocity is
background-dependent, and in principle can be arbitrarily high. This
may be an indication that black holes are bumpy in the DGP model as
well. Unfortunately, an explicit black hole solution which would allow
 the study of perturbations is not yet available in
this model, so one is not able to verify whether this expectation is
true or not.

The rest of the paper is organized as follows. We start with reviewing
the basics of the black hole no-hair theorems in Sect.~\ref{nohair}.
In Sect.~\ref{phases} we review the Lorentz-violating models of
massive gravity. In Sect.~\ref{bh} we describe the spherically
symmetric black hole solution in these theories and some properties of
the rotating black holes.  In Sect.~\ref{hairs} we explain how
instantaneous modes that are generically present in the Higgs sector
of gravity lead to the infinite amount of black hole hairs. To avoid
unnecessary technicalities, instead of massive gravity we consider the
Lorentz violating massive electrodynamics
\cite{Gabadadze:2004iv,Dvali:2005nt}, which is much simpler
technically and shares with the former the relevant physical
properties. We describe this theory in Sect.~\ref{QEDsetup}.  In
Sect.~\ref{grow} we estimate the magnitude of the black hole bumps. In
the minimal models it turns out to be related to the mass of the
gravitational waves; the limits on the latter imply that the bumps are likely to be large
only for the most massive galactic black holes (with masses of order
$10^9M_{\odot}$). We summarize our conclusions in
Sect.~\ref{conclusions}.

\section{No-hair theorems}
\label{nohair}
In order to understand how Lorentz violating models get around the
no-hair theorems let us start with reviewing how they work in the
conventional theories. The aim of this section is to show that in
order to establish the presence of hairs one can simply look for
finite energy solutions of the linearized equations in a black hole
background.

As a simplest example let us consider a scalar field $\phi$ with mass
$m_\phi$.  For simplicity it is convenient to
consider the pre-existing, neutral with respect to the scalar field,
black hole or star and ask what an external observer at the constant
radius $r$ from the object will measure if there is a small amount of
scalar charge falling in.

Clearly, in case of a star an observer will be able to follow what
happens with a charge by accurately measuring the scalar field profile
outside. For the later purposes it is useful to formulate this
somewhat more formally. Namely, the scalar field outside the star (the
quantity which can be measured by the outside observer) satisfies the
source free equation at late times, after the scalar charge crossed
the surface of the star.  The possibility of having a non-trivial
scalar profile is related to the possibility of having a non-vanishing
boundary conditions for a scalar field at the surface of the star,
which encode the information about the fate of the charge inside.
 
The situation is different for a black hole in several
respects. First, as seen by the outside observer, the charge never
crosses the black hole surface, so it appears that the field equation
outside always has sources. Second, there are no signals which can
escape to the outside from inside the horizon, suggesting that the
boundary conditions at the horizon are not capable of ``monitoring"
the inside of the black hole as they do for a star. As we will see
momentarily, due to the large relative redshift between an asymptotic
and a freely falling observer, the first difference is actually a
fake, while the second is important and indeed implies the absence of
hairs.

We proceed by using the ``tortoise" radial coordinate $r$ such that the
$(t r)$ part of the metric is conformally flat, 
\begin{equation}
\label{tortoise}
ds^2=h(r)(dt^2-dr^2)-R(r)^2\l d\theta^2+\sin^2\theta d\varphi^2\r\;,
\end{equation}
where 
\begin{equation}
\label{hr}
h(r)=1-{R_s\over R(r)}\;.
\end{equation}
The explicit relation between the tortoise $r$ and the Schwarzschild
$R$ radial variables is
\begin{equation}
\label{rrstar}
r=R+R_s\log\l {R\over R_s}-1\r\;.
\end{equation}
They coincide far from the black hole at $R\to\infty$, and the
tortoise coordinate $r=-\infty$ at the black hole horizon $R=R_s$.  In
these coordinates the scalar field satisfies the following simple wave
equation,
\begin{equation}
\label{wave}
\left[{\d^2 \over\d t^2}-{\d^2 \over\d r^2}+V(r)\right]\l R\phi\r=j,
\end{equation}
with the potential given by
\begin{equation}
\label{potential}
V(r)=\l 1-{R_s\over R}\r\l{\ell^2\over R^2}+{R_s\over R^3}+m_\phi^2\r\;,
\end{equation}
where $\ell^2$ is the angular operator with eigenvalues $l(l+1)$.  The
form of the source $j$ is determined by the only possible covariant
coupling, $\lambda\int d\tau\phi(x_s(\tau))$, of the scalar field to
the source wordline $x_s(\tau)$ parametrized by the proper time
$\tau$.  As the result one obtains
 \begin{equation}
 \label{coupling}
 j={\lambda\over R \sin{\theta}}{d\tau\over dt}\delta^{(3)}(x^i-x^i_s(t))\;,
 \end{equation}
where $x^i=r,\theta,\varphi$. As the charge approaches the horizon its
proper time changes more and more slowly as seen by the outside
observer, and the source (\ref{coupling}) extinguishes as
\[
{d\tau\over dt}\lesssim\sqrt{1-{R_s\over R}}\;.
\]
We see that, just like for a star, at late times the scalar field
satisfies the source-free equation outside the black hole.

\begin{figure}[t]
\psfrag{V}{$V(r)$}
\psfrag{r}{$r$}
\begin{center}
\epsfig{file=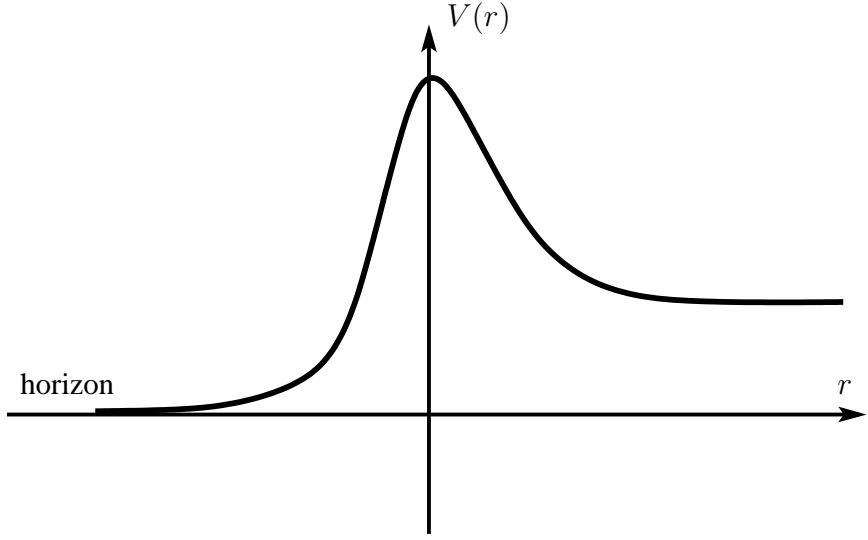,width=0.66\textwidth}
\end{center}
\caption{\small\it The potential for the scalar field
perturbations in the black hole background.}
\label{fig:potential}
\end{figure}

The difference is that the black hole horizon, unlike the star
surface, is at the infinite value $r=-\infty$ of the radial coordinate
$r$ which is a natural one for the scalar field equation
(\ref{wave}). The potential $V(r)$ is shown in
Fig.~\ref{fig:potential}; it is positive everywhere outside the black
hole, and vanishes near the horizon, i.e. at $r\to-\infty$. Clearly,
as expected, in this situation a finite energy charge infalling into
the black hole can source the scalar field outside only for a finite
amount of time. Moreover, the potential (\ref{potential}) does not
allow bound states with finite energy --- all static solutions
decaying at $r= +\infty$ diverge at the black hole horizon.
Consequently, the source-free scalar field perturbations completely
dissolve; they are partially absorbed by the black hole, partially
radiated at infinity, and no hairs remain.  Note that this conclusion
is valid independently of whether the mass of the scalar field
$m_\phi^2$ vanishes or not.

The time scale for the loss of scalar hairs depends on details of the
collapse, and in principle can be arbitrarily long as seen by the
asymptotic observer. Indeed, one can take an initial scalar
perturbation that follow the static solution of the scalar field
equation that decays at $r=+\infty$ all the way until very large
negative values of the radial coordinate $r_{0}$ (meaning very close
to the black hole horizon).  Such a solution will remain unperturbed
in the asymptotic region on the r.h.s. of the potential barrier in
Fig.~\ref{fig:potential} for a time of order $|r_{0}|$. Of course,
this is a very fine-tuned situation; also, at fixed energy the
amplitude of the scalar field goes to zero as $r_0$ grows. More
realistically one expects that the energy of the initial scalar
perturbations is concentrated not too close to the horizon, around
$r\sim R_s$.  Then the time scale for a decay of the scalar hairs is
set by $R_s$.

To summarize, the loss of hairs is a two step process. First, the scalar
field outside the black hole becomes source free. Second, it dissolves
as a consequence of the absence of the static solutions to the
perturbation equation (\ref{wave}). This is similar to how a
perturbation of the sourceless free scalar field would dissolve in the
infinite space.

To illustrate this picture in another example, let us consider a loss
of the massive vector hairs. It is straightforward to check that for
all non-spherical perturbations the situation is identical to that for
the scalar field, up to extra technicalities due to more complicated
tensor structure. One may expect the spherically symmetric case to be
different, because at zero mass of the vector field black hole may
have spherical hairs (electric and magnetic charges), so we
concentrate on this case.

Two non-zero components of a spherically symmetric vector field in the
black hole background (\ref{tortoise}) are $A_0(t,r)$ and
$A_r(t,r)$. They satisfy the usual Maxwell--Proca equations
\begin{equation}
\label{Proca}
{1\over \sqrt{-g}}\d_\mu\l\sqrt{-g} g^{\mu\alpha}g^{\nu\beta}
F_{\alpha\beta}\r+m_A^2g^{\nu\alpha}A_\alpha=j^\nu\;,
\end{equation}
where the electromagnetic current $j^\nu$ has the following form for a
point source,
\[
j^\nu={e\over\sqrt{-g}}\delta^{(3)}(x^i-x_s^i(t)){dx^\mu\over dx^0}\;.
\]
The only non-vanishing component of the electromagnetic strength in
the spherically symmetric case is the radial electric field
$E=F_{tr}$. Taking the divergence of the Proca equations (\ref{Proca})
one obtains the constraint equation \be
\label{constraint}
\d_tA_0={1\over R^2}\d_r\l R^2A_r\r\;.
\ee
Using this equation one can eliminate $A_0$ from the $(r)$ component of the 
Proca equations and arrive at the following equation for $A_r$ alone, 
\be
\label{Areq}
\left[{\d^2 \over\d t^2}-{\d^2 \over\d r^2}+V_A(r)\right]\l RA_r\r
=e\dot{r}\l{1-{R_s\over R}}\r^{1/2}{\delta^{(3)}
(x^i-x_s^i(t))\over R\sin\theta}\;,
\ee
where the potential is
 \[
 V_A=\l1-{R_s\over R}\r\l{2\over R^3}-{6M\over R^4}+{m^2_A\over R}\r\;.
 \]
Just like in the scalar case this equation becomes source-free as the
charge approaches the black hole horizon. It is straightforward to
check that the same is true for a mode with $A_r=0$ and $A_0=A_0(r)$
--- the only one where the value of $A_0$ is not determined from the
constraint equation (\ref{constraint}).
 
So, as before, we just need to check that the massive Proca equations
in the black hole background do not admit static finite energy
solutions without sources. The Proca constraint (\ref{constraint})
implies that $A_r\propto 1/R^2$ for a static solution, however this
does not solve the wave equation (\ref{Areq}) at non-zero mass. Also
the $(t)$ component of the Proca equations implies that $A_0(r)$
satisfies the time-independent Schr\"odinger equation with the positive
potential. This completes the proof of the no-hair theorem for the
massive vector field.
 
It is instructive to see how the massless limit is recovered. To keep
this limit smooth we impose (\ref{constraint}) as a gauge fixing
condition.  Note that this does not fix the gauge freedom completely;
the residual gauge transformations are of the form $A_\mu\to
A_\mu+\d_\mu\alpha$, where $\alpha$ satisfies
\begin{equation}
 \label{residual}
 \d_t^2\alpha-{1\over R^2}\d_r\l R^2\d_r\alpha\r=0.
\end{equation}
In the massless case the $(t)$ component of the Proca equations
reduces to the Gauss law giving the following static electric field as
a solution
\[
F_{tr}\propto {1-R_s/R\over R^2}\;.
\]
The most general vector potential giving rise to this field strength
and satisfying (\ref{constraint}) has the form
\begin{equation}
 \label{procast}
 A_\mu\propto\l -\int_r^\infty dr'{1-R_s/R(r')
\over R(r')^2},\;0\r+\d_\mu\alpha\;,
\end{equation}
where $\alpha$ solves (\ref{residual}). If $\alpha=0$ then $A_\mu^2$
diverges at the horizon and, consequently, in this case
(\ref{procast}) does not correspond to a limit of a smooth family of
solutions of the massive equations. By taking a non-trivial $\alpha$
one can avoid this problem and make $A_\mu^2$ finite at the horizon,
but this inevitably leads to the non-vanishing $A_r$ component
(provided one keeps $A_\mu^2$ zero at the spatial infinity). The $(r)$ component of the Proca equations implies then that at
the non-zero photon mass there is a non-trivial time-dependence as
well.  This time-dependence makes a solution to dissolve on a time
scale which becomes infinite as the mass is sent to zero, $m_A\to 0$ (cf. \cite{Pawl:2004bx}).
 
Note that contrary to what the vector field example seems to suggest,
the no-hair theorems {\it do not} imply that black holes are not able
to support the non-trivial profile for massive fields. For instance, a
massive dilaton coupled to the photon as $\phi F_{\mu\nu}^2$ would
have a non-trivial profile outside a charged black hole (similarly, a
coupling of the type $\phi R_{\mu\nu\lambda\rho}^2$ would generate a
scalar profile outside a neutral black hole as well). The actual
content of these theorems is that the possible continuous deformations
of the black hole metric in the conventional theories are related to
the gauge charges, and all other details of the metric are completely
determined by the latter.

To conclude, we have argued that a natural way to check the existence
of hairs is to check whether the linearized field equations in the
black hole background admit static finite energy solutions. We will
use this procedure in the subsequent sections.

\section{Higgs phases of gravity}
\label{phases}

As explained in the Introduction, we expect the presence of black hole
hairs to be a generic (probably unavoidable) feature of theories with
the spontaneous Lorentz violation.  Analysis of the effective field
theories describing gravity in the Higgs phase reveals a natural
mechanism for generating the hairs: generically, the Goldstone sector
of the consistent Lorentz violating models contains fields mediating
instantaneous interactions. The purpose of this section is to explain
the origin of these fields, and to briefly review the phenomenological
constraints on these models. We also describe the exact black hole
solutions relevant for the later discussion.

\subsection{Setup and its basic properties}
\label{setup}

The idea behind models describing gravity in the Higgs phase is to
introduce a spontaneous breaking of Lorentz invariance induced by the
space-time dependent condensates of the scalar fields. In general one
has four scalar fields $\phi^0$, $\phi^i$ ($i=1,2,3$) with the
following vacuum expectation values (vevs) in the ground state,
\begin{gather}
\label{vevs}
\phi^0=t\;\;,\;
\phi^i=x^i\;.
\end{gather}
In order to preserve the invariance under the space-time translations
one assumes that the scalar fields have purely derivative
couplings. Then the action is invariant under both space-time
translations and shifts of the scalar fields. The scalar vevs in
equation (\ref{vevs}) break both of these symmetries, however a
residual symmetry remains that is the translation accompanied by the
compensating shift in the fields.  As a result, equations
describing dynamics of  perturbations around the ground state
(\ref{vevs}) are invariant under the space-time
translations. Similarly, we assume that the action for the scalar
fields has a global $O(3)$ symmetry under which $\phi^i$'s transform
as a vector. The ground state (\ref{vevs}) is invariant under rotation
of the spatial coordinates accompanied by the global rotation of
$\phi^i$ in the opposite direction, implying rotational symmetry for
perturbations. The covariant action of the theory takes the form
\begin{equation}
\label{verygeneralaction}
\int d^4x\sqrt{-g}\l
M_{Pl}^2R[g]+\Lambda^4F(\d_\mu\phi^0,\d_\mu\phi^i,\dots)\r\;, 
\end{equation}
where $\Lambda$ is a UV cutoff scale and dots stand for terms with
larger number of derivatives acting on the scalar fields.

There is a simple physical interpretation of such systems. Namely, let
us assume for a moment that the function $F$ does not dependent on the
field $\phi^0$. Then the action (\ref{verygeneralaction}) can be
viewed as a Lagrangian description of the homogeneous and isotropic
relativistic medium (fluid/solid/jelly) coupled to gravity. In this
interpretation fields $\phi^i$ can be considered as defining a map from the
physical space to the comoving fluid space. The ground state in
equation (\ref{vevs}) corresponds to the unperturbed fluid, while 
perturbations of the fields $\phi^i$ describe sound waves
(phonons). The field $\phi^0$ also fits naturally in the fluid
picture. The time dependent vev of the shift-invariant scalar field
indicates the presence of the charge (Bose) condensate. In other
words, the action (\ref{verygeneralaction}) describes dynamics of a
generic relativistic {\it super}\/fluid (supersolid, superjelly)
\cite{Son:2005ak,Dubovsky:2005xd}.

When coupled to gravity, phonons and perturbations of the Bose
condensate mix with the metric perturbations and give a mass to the
gravitons in a way similar to the mixing of the Higgs and gauge boson
perturbations in the conventional Higgs mechanism. This is true, of
course, for an arbitrary fluid. However, ordinary fluids have a
non-vanishing energy-momentum tensor already in the ground state. As a
result, the space-time is curved and the dynamics of the metric
perturbations is affected only at the length scales of the order of
the curvature radius. The key property of the fluid Lagrangians that
give rise to the Higgs phases of gravity is that their energy-momentum
tensor vanishes (or, more generally, has the vacuum form
$T_{\mu\nu}\propto g_{\mu\nu}$) in the ground state. This allows to
change the dynamics of the metric perturbations directly in Minkowski
(or de Sitter) space.

The requirement of the vanishing energy-momentum in the presence of
the fluid and the consistency of the low-energy effective theory (in
particular, absence of ghosts and rapid instabilities) are very
restrictive and allow only a very limited number of possible fluid
actions. The analysis of Ref.~\cite{Dubovsky:2004sg} implies that
some of the phonon modes necessarily have the degenerate dispersion
relations of the form 
\begin{equation}
\label{degenerate}
\omega^2=0\; {\mbox{ or }}k_i^2=0\;,
\end{equation}
where $\omega$ and $k$ are frequency and spatial momentum.  Modes with
the $k_i^2=0$ dispersion relation are the instantaneous fields; their
presence is crucial for the existence of the black hole hairs.

In general, the degeneracy of the dispersion relations
(\ref{degenerate}) are broken by the higher-derivative corrections
present in the effective action (\ref{verygeneralaction}). Then the
dispersion relations take the form
\begin{gather}
\label{wk}
\omega^2=a{k^4\over \Lambda^2}\;,\\
k^2=b{\omega^4\over \Lambda^2}\;,
\label{kw}
\end{gather}
where $a$ and $b$ are dimensionless coefficients of order one.  The
dispersion relation (\ref{wk}) is perfectly stable. On the other hand,
the dispersion relation (\ref{kw}) implies that the kinetic term of
the corresponding mode is higher derivative in time.  This inevitably
leads to the catastrophic classical and ghost instabilities within the
domain of validity of the effective theory. To exclude these
instabilities one has to impose symmetries ensuring that the modes
obeying $k^2=0$ do not acquire time kinetic term to all orders in the
derivative expansion and thus remain truly instantaneous. A minimal
symmetry achieving this goal is
\begin{equation}
\label{xisym}
\phi^i\to\phi^i+\xi^i(\phi^0)
\end{equation}
for arbitrary functions $\xi^i$.  As a consequence of this symmetry,
at the one-derivative level the function $F$ in the action
(\ref{verygeneralaction}) takes the form
\[
F=F(X,W^{ij})\;,
\]
where
\begin{gather}
X=g^{\mu\nu}\d_\mu\phi^0\d_\nu\phi^0,\\
W^{ij}=G^{\mu\nu}\d_\mu\phi^i\d_\nu\phi^j,
\label{X}
\end{gather}
and the ``effective metric'' $G^{\mu\nu}$ is given by
\begin{equation}
\label{crazyG}
G^{\mu\nu}=g^{\mu\nu}-{\d^\mu\phi^0\d^\nu\phi^0\over X} 
\end{equation}
In this form the origin of the instantaneous modes is very explicit;
the effective metric (\ref{crazyG}), which determines the propagation
of the $\phi^i$ excitations, is degenerate in the time-like direction
$\d_\mu\phi^0$. As a result, interactions mediated by the $\phi^i$
fields are instantaneous along the space-like surfaces of constant
$\phi^0$. Excitations of $\phi^0$ itself have a dispersion relation
(\ref{wk}).

As shown in \cite{Dubovsky:2005dw}, the cosmological evolution in
these models drives them into the region where an extra symmetry
emerges that has the form
\begin{gather}
\label{lamsym}
\phi^0\to\lambda\phi^0\;,\;\;
\phi^i\to\lambda^{-\gamma}\phi^i
\end{gather}
with $\gamma$ being some fixed real number. One may thus impose this
symmetry from the beginning. Another reason to impose this symmetry
is that, as discussed later, it implies that the linearized scalar
metric perturbations are described by the same equations as in GR ---
in particular, there is no modification to the Newton's law.  In what
follows we assume that this symmetry is exact, so that at the
one-derivative level the function $F$ depends on a single combination
of the scalar fields,
\begin{equation}
\label{F}
F=F(Z^{ij}),
\end{equation}
where  
\[
Z^{ij}=X\l W^{ij}\r^{1/\gamma}.
\]
The simplest consistent Higgs phase of gravity --- ``ghost
condensate" --- corresponds to the case when the action is independent
of the fields $\phi^i$ (loosely speaking, this corresponds to the limit
$\gamma\to\infty$), so that 
\begin{equation}
\label{ghc}
F=F(X)\;.
\end{equation}
In this case the instantaneous interactions are absent and the
excitations of $\phi^0$ have the dispersion relation of the form
(\ref{wk}).

The symmetries (\ref{xisym}), (\ref{lamsym}) may appear somewhat
unusual. To understand better their meaning and to
see how restrictive our choice of the Lorentz breaking sector
(\ref{verygeneralaction}) is, let us note that in the
reparametrization invariant action (\ref{verygeneralaction}) one can
fix the ``unitary'' gauge in which the scalar fields are equal to
their background values (\ref{vevs}). In this gauge the second term in
the action (\ref{verygeneralaction}) takes form of the ``potential"
for the metric components, 
\begin{equation}
 \label{unitarygen}
\int d^4x\sqrt{-g}\l
M_{Pl}^2R[g]+\Lambda^4F(g_{\mu\nu},\d_\lambda g_{\mu\nu},\dots)\r\;, 
\end{equation}
This action explicitly breaks the diffeomorphism invariance of the
theory.  However, the symmetries (\ref{xisym}) and (\ref{lamsym})
imply that in the unitary gauge the action is still invariant under
the subgroup of the full diffeomorphisms generated by the
time-dependent shifts of the spatial coordinates,
\begin{equation}
\label{xisymc}
x^i\to x^i+\xi^i(t)
\end{equation}
and dilatations
\begin{gather}
\label{lamsymc}
t\to\lambda t\;,\;\;
x^i\to\lambda^{-\gamma} x^i.
\end{gather}
As explained in \cite{Dubovsky:2004sg}, an arbitrary action of the
type (\ref{unitarygen}) can be presented in the form
(\ref{verygeneralaction}). Consequently, our choice of the Lorentz
symmetry breaking sector can be considered as a minimal one, when one
does not add new degrees of freedom beyond those already present in
the metric.
 
 \subsection{Phenomenology of massive gravitation}
 \label{phen}

Here we review the basic phenomenological constraints on the model of 
Ref.~\cite{Dubovsky:2004ud,Dubovsky:2005dw}. 

\begin{figure}[t]
\psfrag{V}{$f(Z)$}
\psfrag{r}{$Z$}
\psfrag{z}{\hspace{-10pt}$Z=1$}
\begin{center}
\epsfig{file=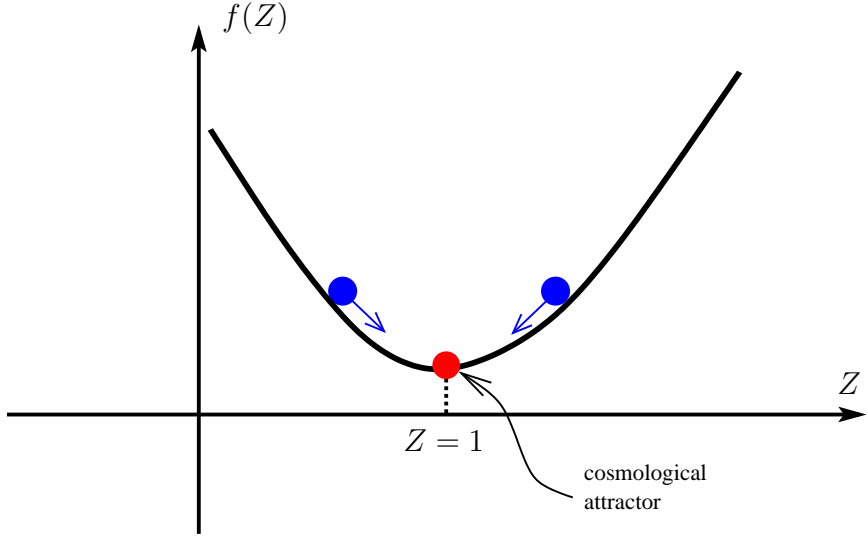,width=0.66\textwidth}
\end{center}
\caption{\small \it The shape of the kinetic function in massive gravity.}
\label{Ffig}
\end{figure}

In order for the energy-momentum tensor to have the vacuum form in the
state (\ref{vevs}) the function
\[
f(Z)=F(Z\delta^{ij})
\]
should have a minimum at $Z=1$ (see Fig. \ref{Ffig}).
The value of the
cosmological constant is determined by the value $f(1)$ at the
minimum. What makes the state (\ref{vevs}) the true ground state (at
least locally in the field space) is that the cosmological expansion
drives $Z$ to the minimum of $f(Z)$.  As one approaches the attractor
point $Z=1$, the cosmological evolution is described by the
conventional Friedmann equation with an extra ``dark" component,
parametrizing deviation from the attractor point.  The equation of
state of this new component is $p=w \rho$ with
\[
w=-{1\over 3\gamma}\;.
\]
This component has a negative pressure for $1/3\leq\gamma\leq\infty$
and a positive pressure for $\gamma<0$. At $0\leq\gamma\leq 1/3$ the
pressure is so negative that the null energy condition is violated. In
this range of $\gamma$ rapid instabilities are present and the
effective field theory is not well-behaved.

The value $\gamma=1/3$ is special. At this point the extra
component behaves as the cosmological constant. What happens is that
at $\gamma=1/3$ the value of $Z$ remains constant during the
cosmological expansion, so there is no need for a function $f(Z)$ to
have a minimum. In this case the observed value of the cosmological
constant is a dynamical quantity determined by the initial conditions
rather than by the parameters of the action.

Let us now describe the behavior of the linearized perturbations
around the Minkowski vacuum in the Higgs phase.  The straightforward
way to do that is to work in the unitary gauge where the scalar fields
are unperturbed and a general metric perturbation takes the form
\begin{eqnarray}
\nonumber
\delta g_{00} &=& 2\varphi,\\
\nonumber
\delta g_{0i} &=&  S_i - \d_i B,\\
\nonumber
\delta g_{ij} &=& - h_{ij} - \d_i F_j - \d_j F_i 
+ 2 (\psi \delta_{ij} - \d_i\d_j E), 
\end{eqnarray}
where $h_{ij}$ are the transverse and traceless tensor perturbations,
$S_i$ and $F_i$ are the transverse vector perturbations, while
$\varphi$, $\psi$, $B$ and $E$ are the scalar perturbations. A
straightforward calculation gives that for an arbitrary
energy-momentum tensor the gauge-invariant scalar potentials $\psi$
and $\Phi\equiv\varphi+\d_0B-\d_0^2E$ are the same as in the general
relativity. Similarly, for an arbitrary source the only
frame-independent combination of the vector perturbations
$(S_i+\d_0F_i)$ is also the same as in the general relativity.  In
fact, this similarity with the general relativity persists in the case
of the expanding Universe where cosmological perturbations behave in
the standard way at least for some values of $\gamma$
\cite{Bebronne:2007qh}.

On the contrary, the transverse traceless metric perturbations
$h_{ij}$ satisfy the {\it massive} Klein--Gordon equation
\[
\l\Box+m_g^2\r h_{ij}=T_{ij}^{TT}\;,
\]
with the graviton mass of the form
\[
m_g^2=\alpha{\Lambda^4\over M_{Pl}^2},
\]
where $\alpha$ is a numerical coefficient of order one which is equal
to a certain combination of  second derivatives of the function $F$
at $Z=1$.

To summarize, the point important for the purposes of the current
paper is that the cosmological expansion and dynamics of the
linearized metric perturbations in the Higgs phases of gravity characterized by the residual symmetries
(\ref{xisymc}), (\ref{lamsymc}) is to a
large extent the same as in the Einstein theory. This opens up a
possibility for a graviton mass to be large, much larger than the
present Hubble constant. The bounds on the graviton mass come
from the properties of the tensor modes. The observations of the
binary pulsars constrain the graviton mass to be smaller than the
typical frequency of the emitted radiation, 
\begin{equation}
\label{masslimit}
m_g\lesssim 10^{-15}\mbox{cm}^{-1}.
\end{equation}
The possibility for gravity waves to have a large mass (compared to
cosmological scales) leads to a number of interesting predictions for
the gravitational wave detectors. First, the graviton mass can be
detected by observing a time delay between the optical and
gravitational wave signal from a distant source. Second, the
non-relativistic massive gravitons can be efficiently produced in the
early Universe (similarly to other light scalar fields such as the
axion). This would lead to a strong monochromatic line with the
frequency set by the graviton mass in the stochastic gravitational
wave background. This signal can be especially strong if the graviton
mass is larger than $\sim 1$~pc$^{-1}$, so that gravitons may cluster
in the Galactic halo. Interestingly, the existing limits do not
exclude that all of the cold dark matter is comprised of the massive
gravitons.  The relevant range of frequencies is covered by the
millisecond pulsar measurements and by LISA, so this possibility has
good chances of being tested in the near future.

There is a more direct effect of the presence of a preferred frame
(the one where the scalar vevs take the form (\ref{vevs})). Namely,
the gravitational field of sources moving relative to this frame is
different from that of static ones. Although the scalar modes are
unaffected by the mass term, the gravitational field of a moving
source has a tensor component. The non-zero mass term for the tensor
modes implies that the predicted gravitational field will be different
from GR. However, in practice this is an extremely small effect as the
tensor component is proportional to the square of the velocity of the
source relative to the preferred frame. Moreover, the effect of this
component on a test mass is also proportional to the square of
velocity of a test particle.  At present these effects are too small
to be observed.

\section{Black Hole Backgrounds}
\label{bh}
As discussed in Sect.~\ref{nohair}, the most straightforward way to
check whether the black hole hairs are present is to start with a known
black hole solution and to check whether the spectrum of its linear
perturbations contains static finite energy modes. Of course, to
implement this program one has to find an explicit black hole solution
first, so the main purpose of the current section is to describe the
simplest black hole solutions in massive gravity. Doing this we also
review some basic facts about the non-linear dynamics in these theories.

\subsection{Ghost condensate}
\label{ghostbh}
The non-linear dynamics of gravity in the Higgs phase is rather
involved already in the simplest case of the ghost condensate model
where the function $F$ depends only on $X$.  An important thing to
keep in mind is that non-linear effects may not be negligible even in
the regime when deviations of the metric from the Minkowski one are
very small~\cite{Arkani-Hamed:2005gu}. One way to understand this is
to note that the field equations of the ghost condensate are
equivalent to the hydrodynamical equations describing irrotational
relativistic fluid with the four-velocity
\[
u_\mu={\d_\mu\phi^0\over\sqrt{X}}. 
\]
It is clear, therefore, that in the presence of massive objects the
equations for the ghost condensate become non-linear at the time scale
of order of the gravitational infall time on these objects. Naively,
this makes it very hard to understand the dynamics of the
model. However, the following observation leads to a radical
simplification in many cases: for an arbitrary metric there
exists (at least locally)  a configuration of the field $\phi^0$ such that 
\begin{equation}
X=1.
\label{eq:X=1}
\end{equation}
The field configuration that satisfies eq.~(\ref{eq:X=1}) is special
in many respects. First, the energy-momentum tensor of the ghost
condensate $\phi$ vanishes at this configuration, so that there is no
gravitational backreaction. Second, $X=1$ is an attractor point of the
cosmological evolution, so for many purposes $X=1$ can be taken as a
natural initial condition. Finally, if one starts with $X=1$ this
relation will continue to hold at least for some finite amount of
time.

A way to understand the latter statement is to note that in the
unitary gauge $\phi^0=t$ the action of the ghost condensate model 
takes the form
\begin{equation}
\label{ghostunitary}
M_{Pl}^2\int d^4x\sqrt{-g}R+\Lambda^4\int d^4x\sqrt{-g}F(g^{00}).
\end{equation}
In the case of the function $F$ having a minimum at $g^{00}=1$ as
shown in Fig.~\ref{Ffig}, this action can be viewed as the Einstein
action with the extra term which fixes the gauge $g^{00}=1$.
Consequently, if one starts with the initial condition satisfying
$X=1$, the solution to the equations which follow from the action
(\ref{ghostunitary}) will coincide with the solution to the {\em
conventional} Einstein equations with the same initial condition,
transformed into the gauge $g^{00}=1$. This makes it clear that a
large part of the non-linear dynamics of the ghost condensate is
related to a question of how to transform a given solution of the
Einstein equations into the gauge $g^{00}=1$, and is irrelevant for an
observer who is not directly coupled to the ghost condensate.

Apart from the situation when one initially starts with $X\neq 1$,
this argument may break for the following reasons. First, the full
ghost condensate action depends not only on $X$, but in general
contains also higher-derivative scalar quantities such as
$(\Box\phi^0)^2$, which are suppressed by the powers of $\Lambda$. With
these terms taken into account, the ghost condensate action in the
unitary gauge does not have the form of the gauge-fixed Einstein
action, and the condition $X=1$ is not preserved during the time
evolution. This leads, for instance, to a slow accretion of the ghost
condensate on black holes as described below.  The limit on the
graviton mass, Eq.~(\ref{masslimit}), implies that
$\Lambda\lesssim 10$~keV. Then these effects are very slow and can be
neglected on time scales of order the current age of the Universe
\cite{Arkani-Hamed:2003uy,Dubovsky:2004qe,Peloso:2004ut,Mukohyama:2005rw}.

Second, for a given solution of the Einstein equations it is not
possible in general to find a globally well-defined transformation to
the gauge $g^{00}=1$. In the fluid language this means that one
expects the caustics to develop in the fluid flow where the fluid
description breaks down. Accounting for these caustics leads to the
space-time pictured as a patchwork of the $X=1$ domains separated by
the caustic regions where the fluid singularities are presumably
resolved in the UV completed theory.

The above arguments suggest that in order to understand the dynamics
of the ghost condensate (at least on reasonably short time-scales and
in sufficiently small regions of space) it suffices to solve the
equation $X=1$ in a fixed Einstein geometry. For instance, a solution
to this equation in the background Schwarzschild geometry takes the
following form (in the Schwarzschild frame) \cite{Mukohyama:2005rw}
\begin{equation}
\label{Schwphi}
\phi^0=t+f(R),
\end{equation}
where
\[
f(R)= 2\sqrt{R R_s}+R_s\ln\l 
{\sqrt{R}-\sqrt{R_s}\over \sqrt{R}+\sqrt{R_s}}\r.
\]
Since the condition $X=1$ implies that the energy-momentum of the
field configuration (\ref{Schwphi}) vanishes, this configuration
together with the Schwarzschild metric solve the equations of motion
which follow from the action (\ref{ghostunitary}). In agreement with the
above discussion, the scalar part (\ref{Schwphi}) of this solution
coincides with the time redefinition that transforms the Schwarzschild
metric into the gauge $g^{00}=1$.

It is straightforward to generalize this solution to the rotating case
and find the Kerr black hole solution in the ghost condensate. The
metric part of this solution is again the usual Kerr metric, and the
ghost condensate field in the Boyer--Lindquist frame has the same
general form (\ref{Schwphi}) with a different function $f(r)$. One may
be surprised that it is possible to find $\phi^0$ which is independent
of the angular variables in the rotating case. Note, however, that the
equation $X=1$ which one needs to solve in order to get zero
energy-momentum of the ghost condensate is just the Hamilton--Jacobi
equation in the Kerr background. This equation is well known, it
allows for separation of variables and as a result one is able to
obtain a solution in the form (\ref{Schwphi}). The explicit form of
the function $f(r)$ is somewhat more complicated in the rotating case,
and we will not present it here.

\subsection{Non-linearities and the simplest black 
holes in massive gravity}
\label{massivebh}

The situation is more complicated in models with gravity in the Higgs
phase and massive graviton. These models possess 4 scalar fields
$\phi^0$ and $\phi^i$ which take non-zero expectation values. In order
for the Schwarzschild and/or Kerr metric to be a solution of the
Einstein equations these condensates have to be such that their
energy-momentum tensor vanishes in the exterior of the black hole. It
is clear from the above discussion that this is possible only when the
condition analogous to eq.~(\ref{eq:X=1}) is satisfied, which has the
form
\begin{equation}
\label{eq:Zdelta}
Z^{ij}=\delta^{ij}.
\end{equation}
As has been explained in Sect.~\ref{phen}, there is no gravitational
backreaction at this point and it is an attractor of the cosmological
evolution, in full similarity with the ghost condensate model.

An important difference between eq.~(\ref{eq:Zdelta}) and its analog
in the case of the ghost condensate (\ref{eq:X=1}) is that for a
generic metric eq.~(\ref{eq:Zdelta}) cannot be solved {\it even
locally}. Indeed, eq.~(\ref{eq:Zdelta}) is a system of six equations
which are, in general, impossible to satisfy with the four fields
$\phi^0$, $\phi^i$.

An equivalent form of eq.~(\ref{eq:Zdelta}) may be obtained if one
goes to the unitary gauge where the action reads
\begin{equation}
\label{massunitary}
M_{Pl}^2\int d^4x\sqrt{-g}R+\Lambda^4\int d^4x
\sqrt{-g}F((g^{00})^\gamma g_{ij}^{-1}).
\end{equation}
The second term in this action does not have a form of the
gauge-fixing term. In order for the contributions of this term to the
field equations to vanish the following six conditions have to be
satisfied,
\begin{equation}
\label{eq:unitary}
(g^{00})^\gamma g_{ij}^{-1}=\delta^{ij}.
\end{equation}
These conditions are, in general, impossible to satisfy with four
coordinate transformations. Note that the counting agrees with the
linear analysis where only two tensor modes acquire a mass and
make the extra term in the action not equivalent to the gauge fixing.

Nevertheless, it is natural to expect that at least for systems
consisting of sufficiently well separated sources with small
quadrupole moments (the latter requirement is not necessary if the
length and time scales involved are all smaller than the inverse mass
of the graviton), the qualitative picture is similar to that in the
ghost condensate --- one obtains a patchwork of domains where
eq.~(\ref{eq:Zdelta}) approximately holds, separated by caustic
regions.

Consequently, it is reasonable to proceed similarly to the case of the
ghost condensate. Namely, given a metric that solves the pure Einstein
equations, one may check whether it is possible to find the scalar
fields such that eq.~(\ref{eq:Zdelta}) is satisfied, so that the
metric is not modified by the backreaction of the condensates.
Eq.~(\ref{eq:unitary}) is an alternative form of
eq.~(\ref{eq:Zdelta}). In geometrical terms, these conditions require,
in particular, that there exists a reference frame such that the
metric induced on spatial slices is conformally flat. A coordinate
transformation to this frame from the original one is determined by
the fields $\phi^0$, $\phi^i$ solving eq.~(\ref{eq:Zdelta}).

A frame where the condition (\ref{eq:unitary}) holds exists for the
Schwarzschild black hole and is called the Gullstrand--Painleve
frame. The black hole metric in this frame is
\begin{equation}
\label{GP}
ds^2=d\tau^2-(dx^i-{R_s^{1/2}\over R^{3/2}} x^id\tau)^2\;,
\end{equation}
where $R=\sqrt{x_1^2+x_2^2+x_3^2}$.  In this frame the scalar field
configuration that solves eq.~(\ref{eq:Zdelta}) is simply
\[
\phi^0=\tau\;,\;\;\phi^i=x^i.
\]
It is straightforward to check that in the Schwarzschild frame the
ghost condensate part of this solution is again given by 
eq.~(\ref{Schwphi}), while the spatial Goldstones are 
$\phi^i=x^i$. Consequently, Lorentz-violating massive gravity
possess a spherically symmetric black hole solution whose
gravitational part is given by the Schwarzschild metric.

Let us now turn to the Kerr metric and see whether there exist
solutions to eqs.~(\ref{eq:unitary}) in that case. Even though the
number of equations is larger than the number of unknowns, it is not
obvious that all the equations are independent in this particular
case. Fortunately, conformally flat spatial slicings are an important
ingredient in the numerical simulations of the black hole mergers, so
their existence for different solutions of the Einstein equations has
been extensively studied~\cite{Garat:2000pn,ValienteKroon:2004gj}. In
particular, it was proven that the conformally-flat slicing of the
Kerr metric is impossible due to the existence of the non-trivial
invariant of the quadrupole origin \cite{ValienteKroon:2004gj}
(loosely speaking, tensor moment)
\begin{equation}
\label{obstruction}
{\Upsilon}=-112\pi J^2.
\end{equation}
Note that this invariant is quadratic in the angular momentum
$J$. Indeed, one may check by direct calculation that
eqs.~(\ref{eq:Zdelta}) for the Kerr metric can be satisfied to the
linear order in $J$. This is a manifestation of the fact that it is
not the angular momentum itself (which is the vector quantity), but
rather a tensor moment that does not allow to satisfy the conditions
(\ref{eq:Zdelta}). This is in accord with the linearized analysis of
massive gravity, where only the tensor part of the metric
perturbations is different from the GR case. Interestingly, the
results of Ref. \cite{ValienteKroon:2004gj} imply that not only the
Kerr metric, but an arbitrary axisymmetric vacuum solution of the
Einstein equations with non-zero angular momentum has a non-vanishing
value of $\Upsilon$ and, consequently, does not allow conformally flat
spatial slicings.

In view of the equivalence between eq.~(\ref{eq:unitary}) and
eq.~(\ref{eq:Zdelta}), the absence of solutions to
eq.~(\ref{eq:unitary}) implies that there do not exist configurations
of the Goldstone fields such that their energy-momentum tensor is zero
in the background of the Kerr metric. Therefore, there are no
solutions in massive gravity that have the Kerr metric as their metric
component.

\section{The toy instantaneous QED model}
\label{QEDsetup}

To understand better the meaning of the above results and to see how
the instantaneous interactions that are present in massive gravity
affect the no-hair theorems, it is useful to consider a simpler
setup. In this section we describe a toy QED
model~\cite{Gabadadze:2004iv,Dvali:2005nt} which shares all the
relevant features of massive gravity.

\subsection{Lorentz violating electrodynamics in flat space}
The flat space action for this model is
\begin{equation}
\label{flatQED}
S=\int d^4 x\l -{1\over 4 e^2}F_{\mu\nu}^2-m^2A_i^2\r.
\end{equation}
Note that the mass term is not the standard Proca term as it only
includes the spatial components of $A_\mu$. Clearly, this mass term
violates Lorentz invariance. To make the analogy with the  massive gravity
more explicit let us perform the scalar/vector decomposition with
respect to spatial rotations. Namely,
if one writes
\[
A_i=\d_i s+a_i\;,
\]
with $a_i$ being the transverse vector, $\d_ia^i=0$, then one obtains
two decoupled sectors, the scalars $(A_0, s)$ and the
vector $a_i$.  The scalar component of the electromagnetic field
induced by an arbitrary distribution of charges is the same as in the
usual electrodynamics in the gauge $s=0$. In particular, the
electrostatic limit in this model is the same as in the usual QED.

On the other hand, the vector perturbations are massive and satisfy
the following equation
\[
\l\Box+m^2\r a_{i}=j_{i}^{T}\;,
\]
where $j_i^{T}$ is a transverse (in the 3-dimensional sense) part of
the electric current. Consequently, the electromagnetic waves acquire
a mass which coexists with the long-range Coulomb potential. Note that
the magnetic field is completely determined by the vector part $a_i$,
and satisfies the usual massive Proca equation
\begin{equation}
\label{magn}
(\Box+m^2)B^i=\epsilon^{ijk}\d_jj_k
\end{equation}
so that no long-range magnetic field is possible.  The
electric field obtains contributions from both the scalar and vector
sectors and satisfies the equation 
\begin{equation}
\label{elec}
(\Box+m^2)E^i=\d_ij_0-\d_0j_i-{m^2\over \d_i^2}\d_ij_0\;.
\end{equation}

An alternative way to understand some of the properties of this model
is to reintroduce gauge invariance by making use of the St\"uckelberg
trick, {\it i.e.} by replacing the vector field $A_\mu$ with the
combination
\begin{equation}
A_\mu\to A_\mu+\d_\mu S,
\label{eq:Stuckelberg-field}
\end{equation}
where $S$ is the St\"uckelberg scalar field. Under a gauge
transformation we now have $A_\mu\to A_\mu+\d_\mu \chi$ and $S\to
S-\chi$, so that the gauge invariance is restored.
 
Since the first term in the action (\ref{flatQED}) is gauge-invariant,
the St\"uckelberg field enters only through the mass of the vector
field. The action (\ref{flatQED}) does not contain a mass term for the
time component $A_0$, and thus the time derivative of the scalar field
$S$ does not appear in the action.  Consequently, unlike in the
conventional massive QED, the field $S$ does not correspond to the new
propagating degree of freedom and the action (\ref{flatQED}) describes
only two propagating modes --- the transverse components of $A_i$ ---
just as in the massless case.  This does not contradict to the
conventional counting of degrees of freedom for Lorentz-invariant
massive vector particles because the action (\ref{flatQED}) does not
possess the symmetry allowing to go into the particle rest frame, which
is necessary for the standard argument. On the other hand, there is no
way to define what ``transverse" means for the zero spatial momentum, so
that a massive photon at rest is characterized by all three spatial
components $A_i$ in agreement with the usual counting.

The St\"uckelberg field enters the action only through its spatial
gradients, so this field can be thought of as a kind of Lagrange
multiplier. Another useful way of thinking about this field is that it
plays a role similar to the electric potential in the electrostatics,
or to the gravitational potential in the Newton's theory of
gravity. This suggests the existence of the instantaneous interactions
in the system (\ref{flatQED}), and indeed the last term on the
r.h.s. of eq. (\ref{elec}) gives rise to the instantaneous electric
field.  This does not lead to problems with causality; the existence
of the preferred reference frame where the action takes the form
(\ref{flatQED}) (in other words, the frame where the time derivatives
of the St\"uckelberg field are absent) allows one to define
unambiguous causal ordering of the events, the time ordering in the
preferred frame.

\subsection{Covariant action}

To make the action (\ref{flatQED}) covariant we need to couple it to a
Higgs sector that spontaneously breaks Lorentz invariance. We choose
the simplest of the models of Sect.~\ref{setup}, namely the ghost
condensate model. Then the covariant form of the action
(\ref{flatQED}) reads
\begin{equation}
\label{ghostQED}
S=\int d^4 x\sqrt{-g}\l F(X)-{1\over 4}
F_{\mu\nu}^2-m^2G_\epsilon^{\mu\nu}A_\mu A_\nu\r.
\end{equation}
Here $X$ is defined in eq.~(\ref{X}) and the function $F$ has the profile
shown in Fig.~\ref{Ffig}.  The spontaneous violation of Lorentz
invariance is mediated to the vector field through the ``effective
metric''
\begin{equation}
 \label{Geff}
 G^{\mu\nu}_\epsilon=-g^{\mu\nu}+\epsilon{\d^\mu\phi^0\d^\nu\phi^0\over X}
\end{equation}
where $\epsilon$ is a parameter varying between 0 and 1. When
$\epsilon=1$ this action reproduces eq.(\ref{flatQED}) in the
Minkowski vacuum $\phi^0=t$. This is the value we are interested in
(cf. effective metric (\ref{crazyG}) for $\phi^i$ fields in massive
gravity).  Note that the choice $\epsilon=1$ is protected by a
residual gauge symmetry with the parameter of the gauge transformation
being constant on the hypersurfaces of constant $\phi^0$,
\[
 A_\mu\to A_\mu+\d_\mu\alpha(\phi^0(x)).
\]
If now we introduce the St\"uckelberg field $S$ according to the
relation (\ref{eq:Stuckelberg-field}), this field will propagate in
the effective metric $G^{\mu\nu}$ just as the fields $\phi^i$ do in
massive gravity. Thus, the Lorentz-violating electrodynamics can be
viewed as a theory of a single instantaneous scalar field whose shift
symmetry is gauged.

\subsection{Black Holes}

Let us discuss what happens with the simplest black hole solutions
in the model (\ref{ghostQED}). Following the logic of Sect.~\ref{bh},
let us check whether the conventional black hole solutions are
preserved in the presence of the Lorentz-violating photon mass.

Clearly, zero charge black hole solutions with $A_\mu=0$ are the same
as in the pure ghost condensate model. In particular, this is the case
for the neutral spherically symmetric black hole solution. As has been
explained in Sect.~\ref{ghostbh}, the metric of this solution is the
usual Schwarzschild metric, while the ghost condensate field in the
Schwarzschild frame takes the form (\ref{Schwphi}).  Similarly, the
neutral Kerr black hole is also a solution in the Lorentz violating
QED.

Let us see now what happens with the charged spherically symmetric
black holes. They have vanishing magnetic field, so it is natural to
expect that it should be possible to find the usual
Reissner--Nordstrom solutions in the Lorentz violating QED as well, as
the electric field remains massless in this case. Indeed, as in
Sect.~\ref{ghostbh}, we can satisfy the equation $X=1$ for the ghost
condensate with the ansatz (\ref{Schwphi}) in the Schwarzschild-like
frame where the Reissner--Nordstrom metric has form
\begin{equation}
\label{RN}
ds^2=h(R)dt^2-h(R)^{-1}dR^2-R^2(d\theta^2+\sin^2\theta d\phi^2).
\end{equation}
The vector field in this frame is equal to
\begin{equation}
\label{ARN}
A_\mu=({Q\over R},0,0,0)+\d_\mu\alpha,
\end{equation}
where $Q$ is the electric charge. Let us show that it is possible to
choose the parameter of the gauge transformation $\alpha$ such that
contributions of the photon mass term to the Maxwell, Einstein and
ghost field equations vanish. The contribution to the Maxwell equations
vanishes if
\begin{equation}
\label{zeroMA}
G_\epsilon^{\mu\nu}A_\nu=0.
\end{equation} 
At $\epsilon=1$ the metric $G_\epsilon^{\mu\nu}$ is
positive-semidefinite with one zero eigenvector $\d_\mu\phi^0$, so to
solve (\ref{zeroMA}) it is enough to find $\alpha$ such that
\begin{equation}
\label{Adphi}
A_\mu=a\d_\mu\phi^0
\end{equation}  
for some function $a$. Making use of eqs.~(\ref{Adphi}) 
and (\ref{Schwphi}) one finds that the gauge parameter
\[
\alpha(R)=\int dR {Q\over R} f'(R)
\]
does the job. It is straightforward to check that for the vector
field of the form (\ref{Adphi}) the contributions of the mass term to the
Einstein and ghost condensate equations of motion vanish as well. So,
in accord with the intuitive expectation the charged spherically
symmetric black hole preserves the Reissner--Nordstrom form in the
Lorentz violating massive QED.

The case of a rotating charged black hole is fundamentally
different. Indeed, the standard Kerr--Newman solution has non-zero
electric and magnetic fields. In particular, its magnetic dipole
moment is equal to
\begin{equation}
\label{mumoment}
\mu=
q J/M\,,
\end{equation}
where $q$ is the electric charge. Lorentz violating massive QED does
not allow long-range magnetic fields, so it does not possess rotating
charged black holes with the same metric and electromagnetic field as
for the Kerr--Newman black hole. A more formal way to say this is to
note that for the Kerr--Newman metric both (pseudo)scalar invariants
of the electromagnetic field, $F^2$ and $F\tilde{F}$, are non-zero. On
the other hand, in order for the mass contribution to the Maxwell
equations to vanish one needs the relation (\ref{Adphi}) to hold. It
is straightforward to check that this relation implies $F\tilde{F}=0$.

We see that the properties of the Lorentz violating QED, and in
particular the fate of the conventional black hole solutions nicely
match with what we found in massive gravity.  A rotating black hole
carries a long-range tensor component of the gravitational field
(quadrupole moment), while a charged rotating black hole possesses a
long-range magnetic field. As a result, these solutions are modified
when the gravitational tensor mode and the magnetic field acquire a
mass. Note that unlike the conventional (Lorentz-invariant) massive
electrodynamics where charged black holes are absent, the black hole
solutions in the Lorentz-breaking models described above should
survive as they are still labeled by the conserved (due to the
residual gauge symmetries) quantities, the angular momentum and
charge. It is only the massive components of the corresponding
solutions which get suppressed far from the black hole (similarly to the
massive dilaton example mentioned in the end of Sect.~\ref{nohair}).

We will discuss phenomenological implications of this result in
Sect.~\ref{grow}.  Now we are ready to address the uniqueness of these
solutions and the fate of the no-hair theorems in the presence of the
instantaneous interactions.

\section{Instantaneous interactions and black hole hairs}
\label{hairs}
It would be really surprising for no-hair theorems to hold in the
presence of instantaneous interactions. Naively, one expects that in
that case the black hole horizon is no longer a special place, and the
information is not lost after the collapse. We will see in this
section that this intuition is perfectly right.  We do not consider
the full theory of massive gravity to avoid unnecessary technical
complications. Instead, we start by analyzing a single instantaneous
scalar field in the black hole background. This example allows us to
understand completely the underlying causal structure.  It makes it
clear that black holes indeed have an infinite amount of hairs
whenever such a field is present. As a concrete example of this
phenomenon in a situation close to one in massive gravity we consider
neutral non-spherically symmetric black holes in the Lorentz violating
electrodynamics of section~\ref{QEDsetup}. In particular, we
demonstrate the existence of hairs which can be interpreted as the
electric dipole moment of a black hole.

\subsection{Instantaneous scalar field}
\label{scalar}
Consider a scalar field $\phi$ interacting with the ghost condensate
and as a result propagating in the effective metric (\ref{Geff}),
\begin{equation}
\label{phi}
S=\int d^4x\sqrt{-g}G^{\mu\nu}_\epsilon\d_\mu\phi\d_\nu\phi\;.
\end{equation}
Eventually, we are interested in the case $\epsilon=1$ when the field
becomes instantaneous, similarly to the Goldstones $\phi^i$ in massive
gravity. However, it is instructive to keep $\epsilon$ general for the
time being.  Just like for $\phi^i$'s the choice $\epsilon=1$ is
protected by a symmetry $\phi\to\phi+\xi(\phi^0)$, where $\xi$ is an
arbitrary function. The field equation following from the action
(\ref{phi}) is
\begin{equation}
\label{coulomb}
{1\over \sqrt{-g}}
\d_\mu\l \sqrt{-g}G_\epsilon^{\mu\nu}\d_\nu \phi\r=j,
\end{equation}
where $j$ is an external current. For $\epsilon<1$ the determinants of
the metrics $g_{\mu\nu}$ and $G_{\epsilon\;\mu\nu}$ are related as
\[
g=(1-\epsilon)G_\epsilon
\]
so that eq.~(\ref{coulomb}) can be rewritten as 
\begin{equation}
\label{dalamb}
\Box_G \phi=j\;, 
\end{equation} 
where $\Box_G$ is the d'Alambertian defined with respect to the metric
$G_\epsilon^{\mu\nu}$.  In the limit $\epsilon=1$ the metric
$G_\epsilon^{\mu\nu}$ becomes degenerate and eq.~(\ref{dalamb})
acquires a simple physical meaning. Namely, on the space-like
hypersurfaces of constant $\phi^0$ the operator $\Box_G$ is just a
3-dimensional Laplacian with respect to the induced
metric. Consequently, at $\epsilon=1$  eq.~(\ref{coulomb}) is saying that on the
hypersurface $\phi^0=\const$ the field $\phi$ coincides with the
instantaneous ``Newton-like" potential induced by the scalar charge
distributed on this surface.
 
\begin{figure}[t]
\psfrag{phi=const}{$(\phi^0=const)$}
\begin{center}
\epsfig{file=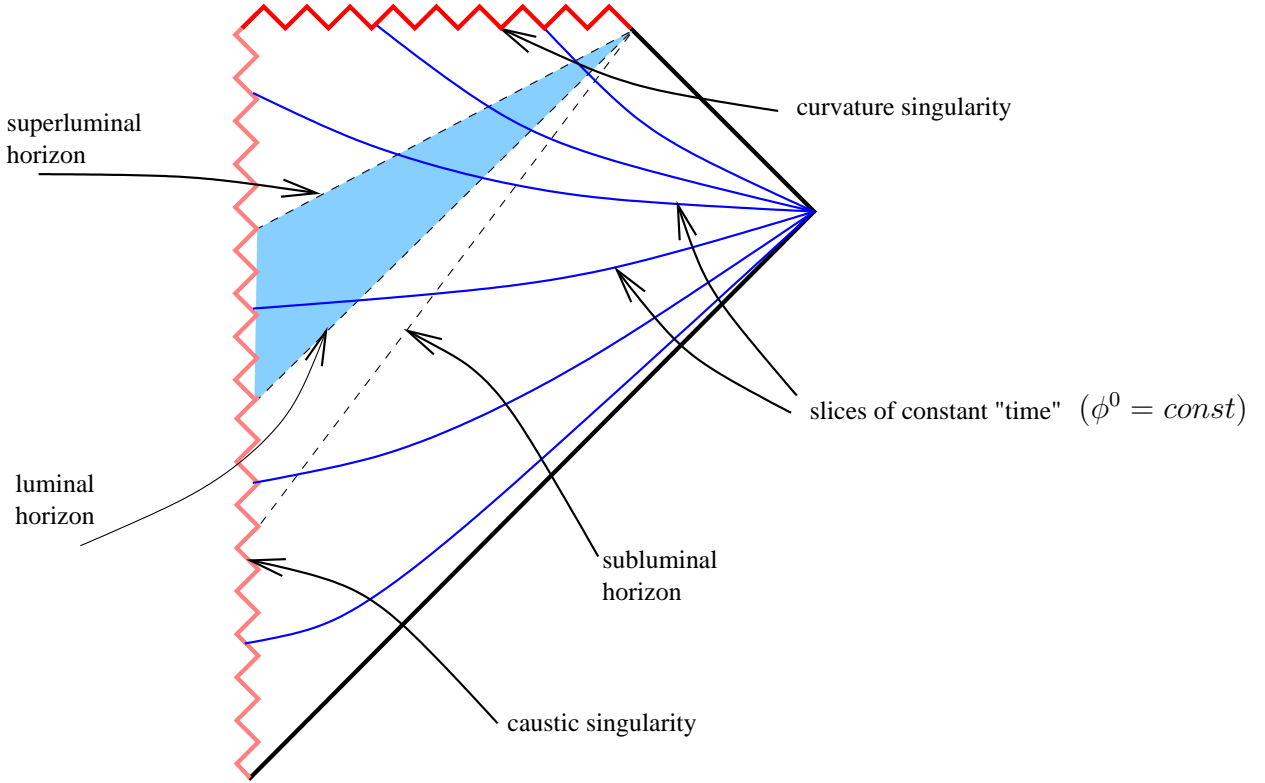,width=0.9\textwidth}
\end{center}
\caption{\small\it Penrose diagram for the black hole formed as a
result of collapse in a Lorentz violating model.  Blue solid lines
correspond to the constant values of the ghost condensate field;
instantaneous interactions make all point on these lines causally
connected to each other. Note, that even before collapse there is a
caustic singularity at the origin (center of the collapsing star).
Dashed lines show the apparent horizons for fields with different
propagation velocities.}
\label{pic:penrose}
\end{figure}

In the flat-space ghost condensate vacuum, the wave equation
(\ref{dalamb}) describes propagation of the field with the sound
velocity
\[
c_\phi^2={1\over 1-\epsilon}\;.
\] 
In the black hole background (\ref{GP}) the time redefinition $\tau\to
\tau(1-\epsilon)^{1/2}$ brings metric $G^{\mu\nu}_\epsilon$ to the
same Gullstrand--Painleve form (\ref{GP}) with a different value of the
Schwarzschild radius
\[
\tilde{R}_s=(1-\epsilon)R_s\;.
\]
Intuitively, this result is very natural: the horizon area is
larger for subluminal fields ($\epsilon<0$) and smaller for
superluminal ones ($\epsilon>0$), see Fig.~\ref{pic:penrose} (the
possibility to look beyond the black hole horizons with a single
derivatively coupled scalar field was discussed recently in
\cite{Babichev:2006vx}).
As $\epsilon\to 1$ the effective Schwarzschild radius $\tilde R_s$
goes to zero, but also the time redefinition becomes singular. At
$\epsilon=1$ the effective metric takes form
\[
G^{\mu\nu}=\l
\begin{array}{cc}
0 & 0\\
0 & -\delta^{ij}
\end{array}
\r\;,
\]
so that (\ref{dalamb}) takes the form of the flat-space Poisson
equation on the surfaces $\tau=const$, in agreement with what we were
saying before. The black hole singularity corresponds to the origin of
the spatial slices. Note that the scalar field equation is not
singular there.
 
We see that already for the superluminal field (for $0<\epsilon<1$)
there is a sense in which black hole may have hairs. Indeed, sources
inside the ordinary $c=1$ horizon, but outside the horizon for the
superluminal field (shaded region in Fig~\ref{pic:penrose}) are
capable of supporting a static non-singular scalar profile outside the
$c=1$ horizon. Of course, these sources themselves should be
superluminal in order to do this as conventional matter necessarily
falls down into the black hole singularity after crossing the $c=1$
horizon. Still this kind of hairs are not enough to resolve the
thermodynamical paradoxes of refs.~\cite{Dubovsky:2006vk,Eling:2007qd}
--- the ``{\em perpetua mobilia}" described there do not require any
matter to be left between the two horizons.
   
In the instantaneous case there is no need for any sources to be
present all the way until the black hole singularity to support a
static scalar hair. Infinitely many kinds of hairs are possible
depending on the source or, equivalently, on the boundary condition at
the singularity. Of course, in general the scalar field constituting
the hairs diverges as its approaches the singularity. But, unlike the
conventional case where it happens at the horizon, in the case of
instantaneous interactions this divergence is localized at the
singularity which is present in any case, so there is no reason to
require the scalar field to be regular there. Eventually, one hopes
that the singularity is resolved in the UV-completed theory. This type
of hairs probe the whole black hole interior and, consequently, are
capable to restore the second law of thermodynamics if they ``grow up"
in the course of the processes described in
\cite{Dubovsky:2006vk,Eling:2007qd}.

Definitely, this sensitivity of the Lorentz violating models to the
boundary conditions at the black hole singularity is not an extremely
appealing property, especially when compared to conventional GR where
cosmic censorship cautiously prevents the asymptotic observers to face
the singularity.  The suggested way to eliminate the thermodynamical
paradoxes may also appear brutal for a person aware of the remarkable
successes of the black hole thermodynamics in GR.  On the other hand,
the sensitivity to the boundary conditions at the short distance
singularities is a rather generic property of the non-linear solutions
in the effective theories, and GR is an exception in this respect. In
addition, it is quite common in gauge theories that the same
phenomenon (information recovery from the black hole in our case)
appears rather differently in the Higgs phase as compared to the
phase with unbroken gauge symmetry.
  
Finally, from the point of view of the external observer who only
measures the field outside of the black hole horizon and does not
directly probe the black hole interior, the Lorentz-violating black
holes are not very different from usual stars.  Indeed, just like in
the conventional case, he sees that sources disappear as they approach
the horizon and then the external field relaxes to some stationary
configuration depending on the details of the collapse.  Such an
observer is not forced to think about singularity, just like for a
star he can assume that the outside field is supported by some smooth
bumpy distribution of matter inside.

At any rate, the above picture seems to be enforced in the Lorentz
violating models by the very general thermodynamical arguments, and
unless it is proven that these models have no chances of being UV
completed, the best one can do is to study where they lead us.
  
\subsection{Lorentz-violating QED}
\label{QED}
There is no conceptual difficulty in extending the above results to
the case when the instantaneous scalar field arises as a Goldstone of
the spontaneously broken gauge or reparametrization symmetry, as it
happens in the Lorentz violating QED and massive gravity.  Let us see
how it works in practice in the technically simpler QED setup. The
Lorentz violating analogue of the Proca equation takes the following
form
\begin{equation}
\label{crazyProca}
{1\over \sqrt{-g}} \d_\mu\l
\sqrt{-g}F^{\mu\nu}\r+m_A^2G^{\mu\nu}A_\mu=j^\nu\;.
\end{equation} 
The black hole hairs are particularly easy to identify in the limit of
the large photon mass.  In the leading order in the $1/m_A$ expansion
the vector field should be of the form $A_\mu=a\d_\mu\phi_0$.  Then
the only equation to be satisfied is a projection of the Proca
equations on the $\d_\mu\phi^0$ direction, which does not contain a
mass term.  As a result one obtains the following equation for the
scalar function $a$,
\begin{equation}
\label{coulomb1}
{1\over \sqrt{-g}}
\d_\mu\l \sqrt{-g}G^{\mu\nu}\d_\nu a\r=\rho
\end{equation}
where $\rho=J^\nu\d_\nu\phi^0$. This equation is identical to the
equation for the instantaneous scalar field of section~\ref{scalar}
and, consequently, in the infinite mass limit Lorentz violating
electrodynamics describes the instantaneous Coulomb force mediated
along the space-like slices of constant $\phi^0$. In the black hole
case these slices are shown in Fig.~\ref{pic:penrose}, and in the same
way as for the scalar field, there is an infinite amount of the black
hole hairs depending on the sources in the black hole interior (in
particular, at the origin of the $\phi^0$ slices --- the black hole
singularity).

At the finite photon mass these hairs get ``dressed" by the
non-vanishing Proca part of the vector field. Let us see how it
happens in a simple example.  Note first that the spherically
symmetric part of the vector field is entirely determined by the
electric charge just like in the massless case.  Indeed, the
spherically symmetric field has only two non-zero components, $A_0$
and $A_r$. In the static case and in the absence of sources the
$r$-component of the modified Proca equations (\ref{crazyProca}) gives
$G^{r\nu}A_\nu=0$. The $(tr)$ part of the effective metric
$G^{\mu\nu}$ is degenerate, so this equation implies also
$G^{t\nu}A_\nu=0$. Hence, in the spherically symmetric case Proca
equations (\ref{crazyProca}) are equivalent to conventional massless
Maxwell equations with the gauge fixing condition
$A_\mu=a\d_\mu\phi_0$. Consequently, there are no charged spherically
symmetric black holes that are different from the conventional
Reissner--Nordstrom ones. In other words, there are no hairs in the
spherically-symmetric sector. 

To provide an example of the exotic electromagnetic hairs let us
consider the dipole ($l=1$) perturbations of the vector field around a
Schwarzschild black hole. In order to make this analysis parallel to
the discussion of the no-hair theorems of the section~\ref{nohair},
let us work in the tortoise coordinates.  We are not changing dynamics
of the sources, so that at late stages of the collapse one is again
solving the empty space equations outside the black hole. Let us check
that, unlike the Lorentz invariant case, there is a nonsingular
static solution of the modified Proca equations (\ref{crazyProca}) for
the $l=1$, $m=0$ vector perturbations in the black hole metric
(\ref{tortoise}). We leave the technical details for the
Appendix~\ref{l_1} and just outline here the main steps and results of
the calculation. A static $l=1$, $m=0$ vector field configuration
can be written in the form
\begin{gather}
\label{vectoransatz}
A_0=y(r)Y_2^0, \\
A_i=v(r)V_{2i}^0+w(r)W_{2i}^0+x(r)X_{2i}^0,
\end{gather}
where $i=r,\theta,\phi$, while $Y_l^m$, $V_l^m$, $W_l^m$ and $X_l^m$
are the scalar and three vector spherical harmonics, respectively.
It is immediate to check that $x(r)$ does not mix with other
variables and satisfies the same equation as in the standard Proca
case. Consequently, there are no black hole hairs involving
non-trivial $x(r)$.

There is no need to explicitly solve the equations for the remaining
functions $y(r)$, $v(r)$ and $w(r)$ to demonstrate the existence of a
regular solution; instead, one can simply count modes. Namely, the
equations determining functions $y(r)$, $v(r)$ and $w(r)$ can be
re-written in terms of a single fourth order linear
equation. Therefore, an arbitrary solution is parametrized by four
real parameters. One of these parameters is an overall normalization,
so we are left with three. The equations are easy to solve both in the
asymptotically flat ($r\to+\infty$) and in the near-horizon
($r\to-\infty$) regions. In the asymptotically flat region one finds
two decreasing and two growing solutions; the requirement that the
growing parts are absent leaves only one parameter. Finally, near the
horizon one finds three regular and one singular solution, so one may
use the remaining parameter to obtain the static solution regular at
the horizon and decaying at infinity --- the dipole hair.

To see how this is different from the massless case in which the hairs
are absent, let us see how a similar counting works in that case.  As
in the Lorentz invariant case, to have a smooth massless limit we
impose the analogue of the Proca constraint following from
(\ref{crazyProca}), \be
\label{gc}
\d_\mu\l \sqrt{-g}G^{\mu\nu}A_\nu\r=0\;, \ee as a gauge fixing
condition. Then one again finds four-parameter family of solutions,
and one of the parameters is an overall normalization.  Out of these
solutions one is singular at the infinity, and one at the black hole
horizon. So, naively one is left with two independent static
solutions, the would-be hairs. However, the gauge condition (\ref{gc})
does not fix the gauge completely: there is a residual gauge freedom
resulting in two pure gauge solutions among the perturbations of the
type we are considering. The would-be hairs found above happen to be a
gauge artifact in the massless limit. This shows that there are no physical dipole hairs of
the Schwarzschild black hole in the conventional Einstein--Maxwell
system.

\vskip0.5cm 

To conclude, in this section we demonstrated the mechanism of
generating the black hole hairs in theories with instantaneous
interactions at the linear level in the field perturbations.  In
principle, one can go to higher orders of perturbation theory and
calculate the backreaction of these hairs to the black hole
geometry. This procedure is completely analogous to the perturbative
reconstruction of, for example, the Reissner--Nordstrom metric in the  limit of a small electric charge
 starting from the Schwarzschild metric. As
in the latter case, we see no reasons for such a perturbative
expansion to break down, and consequently we believe that the presence
of the static hairs at the linearized level indicates the existence of
the exact hairy solutions.  Nevertheless, it would be of interest
(both from the theoretical and phenomenological point of views) to
confirm the existence of such solutions by explicit examples, either
analytical or numerical.

\section{Hair nurturing}
\label{grow}
So far we argued that the coexistence of the spontaneous Lorentz
violation and thermodynamics strongly suggests the presence of an
infinite amount of black hole hairs, and identified the source of
hairs in a large class of Lorentz violating models. However, in order
for these hairs to be relevant for the astrophysical observations
there should be a mechanism to create them during the astrophysical
collapse. For instance, the conventional family of charged black hole
solutions is highly unlikely to have any observational significance as
it is close to impossible to imagine how an electrically charged
astrophysical black hole could have been created or could survive as
such.  Let us discuss the possible scenarios of how Lorentz violating
hairs discussed above could have been generated.  To achieve this, one
has to couple an instantaneous field to conventional matter, so that
it can source hairs during the collapse.

In principle, one can introduce a direct coupling of the form 
\begin{equation}
\label{direct}
S_{direct}=\int d^4x\sqrt{-g}G^{\mu\nu}\d_\mu\phi J_\nu\;,
\end{equation}	
where $J_\nu$ is an arbitrary matter current. This form of the
coupling preserves the symmetry $\phi\to\phi+\xi(\phi^0)$.  We will
not study this possibility here and just mention that the most general
dimension three current one can write with the Standard model fermions
is of the form
\[
J^\mu=\Sigma c_i \bar{\psi}\bar{\sigma}^\mu\psi
\]
and the derivative coupling of the type (\ref{direct}) to conventional
Goldstone bosons gives rise to a spin dependent $1/r^3$ force. This is
not the whole story as the Lorentz violating metric $G^{\mu\nu}$ is present
in the coupling (\ref{direct}); this is likely to give rise to the
spin-independent, but velocity-dependent force as well (recall that
there is a preferred reference frame where $\phi^0=t$).

Given that the instantaneous fields appear as the Goldstones of the
spontaneously broken space diffeomorphisms, it is natural to consider
what happens in the massive gravity models of Sect.~\ref{setup} where
the direct interactions of the instantaneous fields with matter are
absent. By analogy with the case of the large photon mass considered
above we expect that in the regime when the size of the black hole is
large compared to the inverse graviton mass, the tensor component of
the black hole metric is absent and the scalar and vector parts are
completely non-universal as they are determined by the details of the
collapse dynamics (or, equivalently, by the boundary conditions at the
black hole singularity).

Note, however, that the binary pulsar bound on the graviton mass
(\ref{masslimit}) implies that this regime is realized only for the
black holes with mass equal $\mbox{(a few)} \times 10^9$ Solar
masses. Such black holes are expected to exist only in the centers of
the largest galaxies; a typical mass range for the galactic black
holes is $10^5 - 10^7$ Solar masses. Additional problem with observing
the multipole moments of these largest black holes is that it requires
detecting gravity waves of low frequencies, i.e. in the range where
the LISA sensitivity is worse.

So, the opposite regime when the size of the black hole is small
compared to the inverse graviton mass appears to be more relevant for
observations.  In this case one expects hairs to be suppressed simply
because the Lorentz violating massive gravity models were designed to
have a smooth massless limit, and the hairs are absent at the zero
graviton mass.  It is instructive to see how this suppression works.
Let us again consider the toy QED model and see how electromagnetic
hairs emerge during the collapse of the bumpy charge distribution into
the preexisting spherically symmetric neutral black hole in the limit
of a small photon mass (for simplicity we assume that the collapsing
distribution has zero total charge).  To this end let us rearrange the
modified Proca equation (\ref{crazyProca}). Namely, let us apply the
covariant derivative $\nabla^\rho$ to  both sides of
Eq.~(\ref{crazyProca}) and antisymmetrize
with respect to $\rho$ and $\nu$. The resulting equation can be
written in the following form: 
\begin{equation}
\label{massaged}
{\cal D}^2F^{\nu\rho} + m_A^2F^{\nu\rho}
-m_A^2\d^{[\rho}\l{\d^{\nu]}\phi^0\d^\mu\phi^0}X^{-1}{A_\mu}\r
= j^{[\rho\nu]}
\end{equation} 
where 
\[
j^{[\rho\nu]}=\nabla^\rho j^\nu-\nabla^\nu j^\rho\;.
\]
In the flat space the first term in (\ref{massaged}) is simply
$\d_\mu^2$, while in the curved background it also contains the
``mass" terms proportional to the curvature. When the inverse photon
mass is large compared to all other scales in the problem one can
suppress the last term on the l.h.s. of eq.~(\ref{massaged}).  Then
the electromagnetic field strength is the same as in the usual Proca
case (actually, in this approximation it is also the same as in the
pure Maxwell theory). In particular, it vanishes outside of the black
hole on the time scale of order $R_s$ after the
collapse. Consequently, at this order one can write
\begin{equation}
\label{A+B}
A_\mu=A_\mu^{Pr}+\d_\mu\alpha\;,
\end{equation} 
where $A_\mu^{Pr}$ solves the conventional Proca equation
(\ref{Proca}) with the same charge distribution and photon mass.  To
find the ``gauge" part $\alpha$ let us plug the vector field in the
form (\ref{A+B}) into the constraint equation (\ref{gc}) of the
Lorentz violating QED. One obtains
\begin{equation}
\label{alphaeq}
{1\over \sqrt{-g}}
\d_\mu\l \sqrt{-g}G^{\mu\nu}\d_\nu \alpha \r
=-{1\over \sqrt{-g}}\d_\mu\l \sqrt{-g}G^{\mu\nu}A^{Pr}_\nu\r.
\end{equation}
The source for $\alpha$ in the r.h.s. of this equation does not vanish
because the constraint equation (\ref{gc}) is different from the
standard Proca constraint satisfied by $A^{Pr}$. It does vanish
outside the black hole as does $A^{Pr}_\mu$. However, the operator
acting on $\alpha$ in eq.~({\ref{alphaeq}) has a familiar
instantaneous form, so it is enough to have a source inside the black
hole to generate a non-zero $\alpha$ outside. This is just the same
mechanism of generating hairs as discussed before. When plugged back
into eq.~(\ref{massaged}) these $\alpha$-hairs source the
electromagnetic tensor outside the black hole as well. Schematically
one can write
\begin{equation}
\label{schematically}
F^{\nu\rho}={m_A^2\over {\cal D}^2+m_A^2}
\d^{[\nu}\phi^0\d^{\rho]}\l\d^\mu\phi^0\d_\mu\alpha\r\;.
\end{equation}
Here one can estimate ${\cal D}^2$ as
\[
{\cal D}^2\sim{1\over l^2}\;,
\]
where $l$ is a typical length/time scale in the problem. At long
scales $l\gg m_A^{-1}$ there is no suppression; this in in agreement
with the flat space analysis \cite{Dvali:2005nt} that turning on a
source at a given point of space gives rise to the electric field
everywhere in space, which after the time $\sim m_A^{-1}$ is
suppressed only by a distance to the source (this geometric
suppression is coming from solving (\ref{alphaeq})). In the context of
the black hole hairs this implies that far away from the black hole
the {\em relative} strength of the non-universal multipoles is
unsuppressed. However, at distances $l$ from the black hole much
shorter than $m_A^{-1}$ there is an extra suppression by a factor of
$(m_Al)^2$. Note that the above argument provides a way to see the
presence of the hairs which is complementary to the direct
calculations of section \ref{QED}.

The same picture is likely to hold for massive gravity as well. For
small black holes we expect a suppression of the black hole hairs by
the factor of $(m_Al)^2$, where $l$ is the distance to the black
hole. It is hard to reliably estimate the amplitude of the hairs
within the effective field theory because it depends on the boundary
conditions at the singularity. It is therefore likely that one would
need the microscopic theory to calculate it quantitatively.

An estimate of the amplitude of the hairs can be provided by the
following argument.  Just before the collapse the metric is that of a
star of the size $\sim R_s$ and of the mass $\sim M_{bh}$.  The
largest value of the $n$-th mass or angular momentum multipole of such
a start is bounded by
\begin{equation}
\label{makes_sense}
M_n,\;J_n\lesssim M_{bh}R_s^n\;.
\end{equation}
Assuming that after the collapse the dynamics inside the black hole
does not significantly change the multipoles of the energy-momentum
tensor, the above argument indicates that at large distances from the
black hole $l\gg m_A^{-1}$ the part of the metric corresponding to
these multipoles will remain unchanged, while at short distances $l\ll
m_A^{-1}$ this part will be radiated away resulting in the suppression
of the hairs by the factor $(lm_A)^2$.
 
This is a natural generalization of the usual bound on the angular
moment of the black hole $J<M_{bh}R_s$. It follows from the intuition
that the multipoles are supported (due to the presence of the 
instantaneous interactions) by the conventional matter of mass
$M_{bh}$ collapsed behind the black hole horizon of size $R_s$. It
would be interesting to check whether this bound is indeed satisfied
for the gravitational collapse in massive gravity.  Note, however,
that higher multipoles do not correspond to any conserved charges, so
in principle the dynamics inside the horizon can violate the above
estimate leading to either the additional suppression of the hairs or 
to larger values of the multipoles. In principle, it is not
clear even that the bound $J<M_{bh}R_s$ for the angular momentum needs
to be satisfied in the Lorentz violating models, especially taking
into account that perturbations around Lorentz violating vacuum may
carry negative gravitational energy. So the verification of this bound
will already be quite a non-trivial check of the GR predictions.
 
If the conservative bound (\ref{makes_sense}) is correct, it makes it
really challenging to observe hairs of the galactic black holes of a
typical size of $10^5 - 10^7M_\odot$ at least in the minimal massive
gravity models described in Sect.~\ref{setup}. Note, however, that the
limitation comes from the bound (\ref{masslimit}) on the mass of the
tensor gravitational waves which happens also to control the size of
the hairs. One may expect that this property is not generic, and hairs
can be prominent in more general models involving, for instance, extra
light fields in the Lorentz-breaking sector. In principle, even in the
minimal model discussed here there is a room for a larger
effect. Indeed, so far we assumed that there is a single mass scale
$\Lambda$ in the symmetry breaking sector (see,
eq.~(\ref{verygeneralaction})). However natural, this assumption could
easily be avoided by tuning the function $F$ in
(\ref{verygeneralaction}) to give graviton a mass which is much
smaller than the natural value $\sim \Lambda^2/M_{Pl}$.  In any case,
one potentially useful lesson from studying the minimal model is that
observations of the largest black holes, and at the largest possible
distance from the black hole are likely to provide the strongest
constraints. This is not surprising given that it is an IR
modification of gravity which gives rise to the black hole hairs.

\section{Concluding remarks}
\label{conclusions}
To summarize, we provided a general thermodynamical argument
indicating that consistent microscopic theories spontaneously breaking
Lorentz invariance, if they exist, are most likely to violate the
black hole no-hair theorems by allowing an {\it infinite} amount of
black hole hairs. In particular, there is no reason in these theories
to expect the higher multipole moments of the black hole metric to be
universal. We identified a mechanism to generate the hairs in a broad
class of theories describing gravity in the Higgs phase, which relies
on the possibility to ``see" inside the black hole horizon due to the
presence of the instantaneous fields in the gravitational Higgs
sector. In the minimal model this effect is likely to be suppressed
for the typical galactic black holes due to the limit on the graviton
mass which is controlled by the same parameter. This suppression may
disappear for black hole masses approaching 10$^9M_\odot$.  It would
be interesting to understand whether this limitation can be avoided in
more general models (e.g. in the bi-metric
models~\cite{Berezhiani:2007zf,Blas:2007ep}).

A peculiarity of the scenario discussed here is that the black hole
metric depends on the boundary condition at the singularity, and thus
is in principle sensitive to quantum gravity effects.  Consequently,
in this class of models black holes literally provide a probe of
quantum gravity.  This is not so unusual for non-linear solutions in
the effective theories, a notable exception being the conventional GR
which provides a mechanism (``cosmic censorship") to prevent large
class of observers from probing the physics at singularities.

As an open question we mention that there is a subtlety in defining
the multipole moments for black holes discussed here. Indeed, the
conventional multipole moments are defined by assuming that the metric
satisfies the vacuum Einstein equations, which is not the case for
rotating black holes in massive gravity. Nevertheless, it is very
likely that the tests of universality of the black hole moments
assuming that the metric satisfies the vacuum Einstein equations (the
conventional procedure) will be sensitive to the non-universality of
the type we discuss as well. Still, it would be useful to develop a
model-independent approach that would not make such an assumption.
  
Finally, note that the instantaneous effects used here to look behind
the black hole horizons are likely to allow to look behind the
cosmological horizon as well. It may be interesting to study the
possible consequences of this effect in more detail.

\section*{Acknowledgments}
We thank Nima Arkani-Hamed, Gia Dvali, Markus Luty, Alberto Nicolis,
Raman Sundrum and Matt Schwartz for interesting and stimulating
discussions. The work of P.T. is supported in part by IISN, Belgian
Science Policy (under contract IAP V/27).

\appendix
\section{UV modifications of gravity and black holes}
\label{UV}
Let us assume that the cutoff scale in the gravitational sector is
$\Lambda$.  Then, in a ``model independent" way one can write the
covariant effective gravitational action at scales below $\Lambda$ as
\begin{equation}
\label{dumb}
S_{gr}=\int d^4x\sqrt{-g}\l
M_{Pl}^2R+a_1{M_{Pl}^2\over\Lambda^2}R^2
+\dots+a_{n-1}{R^n\over\Lambda^{(2n-4)}}\left({M_{Pl}\over\Lambda}\right)^n
+\dots\;,
\r
\end{equation}
where, in general, $a_n$ are of order one. Of course, this action
contains also terms with more covariant derivatives (each covariant
derivative brings an extra factor of $\Lambda^{-1}$), such as
${M_{Pl}^2/\Lambda^4}\nabla_\mu R\nabla^\mu R$.  The leading vertex
from the $n$-th term in eq.~(\ref{dumb}) is $\sim (\partial^2 h)^n$,
so the choice of $\left({M_{Pl}/\Lambda}\right)^n$ in eq.~(\ref{dumb})
ensures that the corresponding amplitude becomes of order one at
energies of order $\Lambda$. This is the largest coefficient one can
put in front of $R^n$ terms while keeping the effective theory valid up to
the scale $\Lambda$.


In this model-independent scenario, there is a limit on $\Lambda$
coming from the short-distance tests of the Newton's law (for
instance, from the modifications of the coupling to matter via
generation of operators like $ \Lambda^{-2}\int
d^4x\sqrt{-g}R_{\mu\nu}T^{\mu\nu}$, where $T_{\mu\nu}$ is the
energy-momentum tensor), so that $\Lambda$, in principle, can be as
low as a fraction of mm$^{-1}$.
Note, however, that these are probes of gravity in the linear regime,
so there is a room for significant classical non-linear effects due to
the high order terms in (\ref{dumb}) at the much longer distance
scales, provided the curvature is large enough.

Namely, the contribution of the $n$-th term in the action (\ref{dumb})
to the classical equations equals to that of the first for curvatures
\[
R\geq{\Lambda^3\over M_{Pl}}\left({M_{Pl}\over\Lambda}\right)^{1/(n-1)}.
\]
Hence the classical contribution grows with $n$ and asymptotically the
critical curvature is
\[
R_{crit}\simeq\Lambda^3/M_{Pl}\simeq (10^{15}\mbox{cm})^{-2}\l \Lambda\cdot\mbox{mm}\r^3
\]
Requiring that these non-linear effects are small at the BBN epoch
would impose slightly stronger constraint on $\Lambda$ than coming
from short distance tests, namely one needs $R_{crit}\gtrsim \l
10^{11}\mbox{cm}\r^{-2}$. In principle, this still leaves a room for
new non-linear effects at the scales of the galactic black holes.
Note, however, that it is just non-linearities themselves, rather than
the presence of horizon, that are needed for these effects to show
up. Also, these effects are much more likely to show up for smaller
scale non-linear systems such as binary neutron stars and stellar
mass black holes.

On the theoretical side it is extremely hard to imagine a viable model
achieving such a low cutoff scale in gravity.  Note that unlike
massive gravity where the low cutoff is present in a separate
gravitationally coupled Higgs sector, here this is gravity itself
which is strongly coupled at the scale $\Lambda$. For instance, in
string theory the cutoff of gravity is at the string scale
$\Lambda\sim\alpha'^{-1/2}$, and above this scale an infinite tower of
new gravitationally interacting particles arises. Typically this
results in too strong constraints on the cutoff scale to allow
non-linear effects at the scale of the galactic black holes (see
Ref.~\cite{Dvali:2001gx} for a discussion of the constraints on the
gravitational cutoff in string-inspired scenarios).

Finally, note that the coefficients $a_i$ in the effective action
(\ref{dumb}) can be parametrically smaller in a particular model.  For
instance, the expansion of the gravitational action following from the
string theory gives
\[
M_{Pl}^2(R+\alpha'R^2+\alpha'^2R^3+\dots)\;.
\]
In this case one need curvatures of order the cutoff scale in order
for non-linear effect to be large.

To summarize, it appears highly non-plausible that UV effects in the
gravitational sector can be important at the scales of the galactic
black holes.  Even if such effects were present, they would have
nothing to do with the presence of the horizon. In particular, there
are no reasons to expect that they would lead to the non-universality
of the black hole multipoles.

\section{Electric dipole hair}
\label{l_1}
Here we provide some details of the calculation that shows the
presence of a static deformation (hair) of the Schwarzschild black
hole with $l=1$, $m=0$ vector field in the Lorentz violating QED.  At
general values of $l$ and $m$ the spherical vector harmonics used in
eq.~(\ref{vectoransatz}) have the following form,
\begin{eqnarray}
V_l^m&=&\l
 0,\; -\sqrt{l+1\over 2l+1}Y_l^m,\; {1\over\sqrt{(l+1)(2l+1)}}r
\d_\theta Y_l^m,\; {i m\over\sqrt{(l+1)(2l+1)}}rY_l^m
\r,\\
W_l^m&=&\l
 0,\; \sqrt{1\over 2l+1}Y_l^m,\; {1\over\sqrt{l(2l+1)}}r\d_\theta Y_l^m,\; {i m\over\sqrt{l(2l+1)}}rY_l^m
\r,\\
X_l^m&=&\l
 0,\; 0,\; -{m\over\sqrt{l(l+1)}\sin{\theta}}r Y_l^m,\; -{i \over\sqrt{l(l+1)}}r\sin{\theta}\d_\theta Y_l^m
\r\;.
\end{eqnarray}
Here $Y_l^m$ are the usual scalar spherical harmonics. For the problem at
hand it is convenient to redefine the coefficient functions $y$, $v$,
$w$ in the following way,
\begin{gather}
\nonumber
y(r)=2\sqrt{\pi\over 3}
\l
\alpha(r)-\sqrt{1-h(r)}\beta(r)
\r,\\
\nonumber
v(r)=\sqrt{2}
\l-2\sqrt{\pi}
\l \sqrt{1-h(r)}\alpha(r)+\beta(r)
\r
+\delta(r)
\r,\\
\nonumber
w(r)={1\over 3}
\l2\sqrt{\pi}\l \sqrt{1-h(r)}\alpha(r)+\beta(r)\r+2\delta(r)\r\;.
\end{gather}
Then at $l=1$, $m=0$ the explicit form of the vector field components
in terms of the functions $\alpha$, $\beta$, $\delta$ and $x$ is
\begin{gather}
\label{As}
A_0= \cos{\theta}\l \alpha(r)-\sqrt{1-h(r)}\beta(r)\r,\\
A_r=\cos{\theta}\l\sqrt{1-h(r)}\alpha(r)+\beta(r)\r,\\
A_\theta=-{r\sin{\theta}\delta(r)\over 2\sqrt{\pi}}\\
A_\phi=-\sqrt{3\over 8\pi}r\sin^2\theta x(r)
\end{gather}
By plugging these expressions into the modified Proca equations
(\ref{Areq}) one gets a set of the ordinary differential equations for
the radial functions $\alpha$, $\beta$, $\delta$ and $x$.  There is no
point in writing down these equations explicitly. Note first that the
equation for the function $x$ decouples from the rest and is the same
as in the conventional Proca theory. So there is no hairs associated
with this function and we set $x=0$ in what follows. Using the
remaining equations one can solve for $\alpha$ and $\delta$ in terms
of $\beta$ and obtain the 4th order differential equation for $\beta$
alone. This equation is rather involved, but as explained in the main
text, all that we are doing is the mode counting, so we need an
explicit form of this equation only in the limits $r\to \pm\infty$.
In the leading order in the near-horizon limit $r\to-\infty$ this
equation takes form 
\begin{equation}
\label{betahorizon}
R_s^4\beta^{(4)}-8R_s^3\beta^{(3)}+23R_s^2\beta''-28R_s\beta'+12\beta=0
\end{equation}
A general solution for $\beta$ in the near horizon limit is 
\[
\beta=C_1\e^{r/R_s}+C_2\e^{3r/R_s}+C_3\e^{2r/R_s}
+C_4\l\e^{2r/R_s}(2R+R_s)+{\cal{O}}(\e^{3r/R_s})\r
+{\cal{O}}(\e^{4r/R_s})
\]
It is important to keep track of the order of magnitude of the
subleading terms as $\alpha$ and $\delta$ are enhanced relative to
$\beta$ by a factor of $\e^{2R/R_s}$. As a result,
eq.~(\ref{betahorizon}) allows to find $A_\mu$ and $F_{\mu\nu}$ near
the horizon including terms of order ${\cal{O}}(1)$ or even
${\cal{O}}(\e^{r/R_s})$ if $C_4=0$. For instance, at the order
${\cal{O}}(1)$ the vector field at the horizon is
\begin{gather}
\label{horizonvect}
A_\mu dx^\mu=R_s\e \cos{\theta}(\e C_4-m_A^2R_sC_1)(dt+dr)
-{R_s\over 2}\e\sin{\theta}\l7C_1+2\e C_3+2\e(r+R_s)C_4\r d\theta\;.
\end{gather}
The most reliable way to check whether this field is regular at the
horizon is to transform it to the Kruskal coordinates (or any other
coordinate system regular at the horizon).  In this way one finds that
the vector field (\ref{horizonvect}) is always regular at the horizon,
while the corresponding field strength is also regular iff $C_4=0$.
Therefore the Lorentz violating Proca equation has three linearly
independent solutions that are regular at the horizon. A nice
consistency check of the calculation is that in the massless limit
$m_A^2=0$ in the order ${\cal{O}}(\e^{r/R_s})$ the field strength
depends only on the two independent combinations of the coefficients
$C_i$, namely on $C_4$ and $(8C_1+2\e^2 C_2+9\e C_3)$. This is in
agreement with the existence of the two pure-gauge solutions
$A_\mu=\d_\mu(a(r)Y_2^0)$ satisfying the gauge condition (\ref{gc}).
Note that this mode counting is different from the usual Proca case
where the same procedure gives only two (instead of three) solutions
regular at the horizon.

A similar analysis reveals two finite-energy solutions at the spatial
infinity, one of them is the electric dipole with $E\sim 1/r^3$ and
another the exponentially damped magnetic dipole with $B\sim \e^{-m_A
r}$. Another pair of solutions has an infinite energy --- for one of
them the amplitude of the electric field approaches a constant, and
for the other the magnetic field grows exponentially.  Consequently,
the mode counting argument implies that there is only one mode which
is regular both at the horizon and at infinity. This mode becomes pure
gauge in the massless limit.

\end{document}